\documentclass[aps,twocolumn,superscriptaddress]{revtex4}
\usepackage[dvips,final]{graphicx}
  \usepackage{amssymb,amsmath,epsfig,bm,pifont}
   \usepackage{extarrows}
   \graphicspath{{./figs/}}

\newcommand{\A}{{\mathcal{A}}}
\newcommand{\M}{{\mathfrak{A}}}
\newcommand{\I}{{\mathcal{I}}}
\usepackage{relsize}
\newcommand{\babar}{{\mbox{\slshape B\kern-0.1em{\smaller A}\kern-0.1em
            B\kern-0.1em{\smaller A\kern-0.2em R}}}
\def\MSbar{\relax\ifmmode\overline                        
            {\rm MS}\else{$\overline{\rm MS}${ }}\fi}     
           }                                              
\def\MSbar{\relax\ifmmode\overline                        
            {\rm MS}\else{$\overline{\rm MS}${ }}\fi}     
\def\1{\hbox{{1}\kern-.25em\hbox{l}}}

 \date{\today}
 \preprint{\hbox{RUB-TPII-01/2021}}
\newcommand{\mycomment}[1]{\iffalse  #1  \fi}                               
\def\Im{\relax{\textbf{Im}{}}}                            
\def\Re{\relax{\textbf{Re}{}}}                            

\newcommand{\be}{\begin{equation}}\newcommand{\ee}{\end{equation}}%
\newcommand{\bd}{\begin{displaymath}}\newcommand{\ed}{\end{displaymath}}
\newcommand{\bit}{\begin{itemize}}                        
 \newcommand{\eit}{\end{itemize}}                         
\newcommand{\ben}{\begin{enumerate}}                      
 \newcommand{\een}{\end{enumerate}}                       
\newcommand{\baa}{\begin{array}{lll}}                     
 \newcommand{\eaa}{\end{array}}                           
\newcommand{\ba}{\begin{eqnarray}}                        
 \newcommand{\ea}{\end{eqnarray}}                         
\newcommand{\la}{\label}                                  
 \newcommand{\nn}{\nonumber}                              
\newcommand{\gev}[1]{\relax\ifmmode{\text{GeV}^{#1}}      
                     \else{GeV$^{#1}${ }}\fi}             
\newcommand{\Gev}{\relax\ifmmode{\text{GeV}}              
                     \else{GeV{ }}\fi}                    
\newcommand{\Mev}{\relax\ifmmode{\text{MeV}}              
                     \else{MeV{ }}\fi}                    
\def\MSbar{\relax\ifmmode\overline                        
            {\rm MS}\else{$\overline{\rm MS}${ }}\fi}     
\def\as{\relax\ifmmode \alpha_s\else{$ \alpha_s${ }}\fi}  
\def\abar{\relax\ifmmode{\bar{a}}\else{$\bar{a}${ }}\fi}  

\usepackage{color}

\definecolor{green}{rgb}{0.133,0.56,0}

\definecolor{DarkGreen}{rgb}{0.04,0.5,0.1}
 
\definecolor{GrayW}{rgb}{0.50196,0.50196,0.50196}
 \newcommand{\GrayW}[1]{{\color{GrayW} #1}}

   \newcommand{\RedTn}[1]{\textcolor{red}{#1}}

\begin{document}
\title{Extending the application of the LCSR method to low momenta
       using QCD renormalization-group summation. 
       Theory and phenomenology}\thanks{This work is dedicated to the memory of Anatoly Vassilievich Efremov, our mentor,
       friend and colleague.}

\author{S.~V.~Mikhailov}
\email{mikhs@theor.jinr.ru}
\affiliation{Bogoliubov Laboratory of Theoretical Physics, JINR,
             141980 Dubna, Russia\\}

\author{A.~V.~Pimikov}
\email{pimikov@mail.ru}
\affiliation{Institute of Modern Physics, Chinese Academy of Sciences,
             Lanzhou, 730000, P. R. China\\}
\affiliation{Bogoliubov Laboratory of Theoretical Physics, JINR,
	         141980 Dubna, Russia\\}
\affiliation{Research Institute of Physics, Southern Federal University,
             Rostov-na-Donu 344090, Russia}

\author{N.~G.~Stefanis}
\email{stefanis@tp2.ruhr-uni-bochum.de}
\affiliation{Ruhr-Universit\"{a}t Bochum,
             Fakult\"{a}t f\"{u}r Physik und Astronomie,
             Institut f\"{u}r Theoretische Physik II,
             D-44780 Bochum, Germany\\}

\date{\today}
\begin{abstract}
We show that using renormalization-group summation to generate the QCD radiative corrections
to the $\pi-\gamma$ transition form factor, calculated with lightcone sum rules (LCSR), renders the
strong coupling free of Landau singularities while preserving the QCD form-factor asymptotics.
This enables a reliable applicability of the LCSR method to momenta well below 1~GeV$^2$.
This way, one can use the new preliminary BESIII data with unprecedented accuracy below
1.5~GeV$^2$ to fine tune the prefactor of the twist-six contribution.
Using a combined fit to all available data below 3.1~GeV$^2$, we are able to determine
all nonperturbative scale parameters and a few Gegenbauer coefficients entering
the calculation of the form factor.
Employing these ingredients, we determine a pion distribution amplitude with conformal
coefficients $(b_2,b_4)$ that agree at the
$1\sigma$ level with the data for $Q^2 \leqslant 3.1$~GeV$^2$ and fulfill at the same time the
lattice constraints on $b_2$ at N$^3$LO together with the constraints from QCD sum rules with nonlocal condensates.
The form-factor prediction calculated herewith reproduces the data below 1~GeV$^2$
significantly better than analogous predictions based on a fixed-order power-series expansion in the
strong coupling constant.
\end{abstract}
\maketitle

\section{Introduction}
\label{sec:intro}
A useful scheme to consider quantitatively exclusive reactions
of hadrons in QCD is provided by the method of lightcone sum rules
(LCSRs) in terms of a dispersion relation
\cite{Balitsky:1989ry,Khodjamirian:1997tk}.
The core advantage of this calculational scheme is that it incorporates
collinear factorization and the operator product expansion (OPE) on
the lightcone.
Especially the pion-photon transition form factor (TFF)
measured in single-tag experiments
has been analyzed extensively within this approach because one can include in the
dispersion relation the physical photon using a vector-meson resonance
in the spectral density.
However, the applicability of LCSRs at $Q^2$ values below the typical
hadronic scale of $\mathcal{O}(1~\text{GeV}^2)$ is limited.
This is related to the fact that one includes QCD radiative corrections
in terms of a power series expansion order by order of the strong
coupling using fixed-order perturbation theory (FOPT).
But the successive inclusion of such terms suffers from a restricted
accuracy, especially at low momenta, because particular terms of the
expansion may give too strong contributions that would eventually
be offset by neglected still higher-order terms.
To make progress, it would be desirable, even necessary, to perform a
summation of such terms using the renormalization group (RG).
This work is devoted to this task and extends further the previous
analysis in \cite{Ayala:2018ifo} (see also \cite{Ayala:2019etj}),
both conceptually and computationally.
The resulting phenomenological improvements are
also worked out.

In essence, the present approach is based on the RG summation of QCD
radiative corrections by combining the formal
solution of the Efremov-Radyushkin-Brodsky-Lepage (ERBL)
\cite{Efremov:1978rn,Lepage:1980fj} evolution equation
with a dispersion relation.
This combination generates a new kind of strong couplings and exceeds
the standard formulation of the LCSRs in the framework of FOPT.
The emerging modified scheme of LCSRs amounts to a particular version
of fractional analytic perturbation theory (FAPT)
\cite{Bakulev:2005gw,Bakulev:2006ex}---FAPT/LCSR.
FAPT extends the original APT, introduced by Shirkov and Solovtsov
\cite{Shirkov:1997wi,Shirkov:2006gv}
for integer powers of the strong coupling, to any real power in both the
Euclidean and the Minkowski space,
see \cite{Bakulev:2008td,Stefanis:2009kv} for reviews and
\cite{Karanikas:2001cs} for paving the way for this development.
The crucial advantage of the FAPT/LCSR scheme is that it ensures the
analyticity of the strong coupling by rearranging the power series
expansion into a nonpower series of FAPT couplings that have
no Landau singularities
when $Q^2 \simeq \Lambda^2_{\rm QCD}$
\cite{Bakulev:2005gw,Bakulev:2006ex}.
However, in order to include the RG summation,
a further generalization of the
FAPT procedure is necessary, as first discussed in \cite{Ayala:2018ifo}.
To this end, a new analytic coupling $\I_{\nu}$ has to be introduced
that generalizes the previous FAPT couplings $\A_{\nu}$, $\M_{\nu}$,
in the Euclidean and Minkowski region, respectively, in the sense that
they now appear as limiting cases of the new coupling
\cite{Ayala:2018ifo,Ayala:2019etj}.
As a  result, the domain of applicability of the QCD perturbative expansion
within the FAPT/LCSR approach is significantly extended towards lower
momentum transfers allowing a comparison with the data
within a more reliable margin of error.

Phenomenologically, this is all the more important in the case
of the preliminary BESIII data
which bear below $Q^2 < 1.5$~GeV$^2$ an
unprecedented accuracy \cite{Redmer:2018uew}.
As shown in \cite{Stefanis:2019cfn,Stefanis:2020rnd}, the LCSR
predictions within FOPT tend to underestimate these low-$Q^2$ data points.
In this work, we derive a TFF prediction within FAPT/LCSR that
provides a significantly better agreement in this low-momentum regime.
To achieve this goal, we perform a fine tuning of the
nonperturbative scale factors
$\delta^2_\text{tw-4}$ (twist four) and $\delta^2_\text{tw-6}$	(twist six)
with the help of a confection of data from different experiments
in the momentum interval $Q^2\leqslant 3.1$~GeV$^2$.
We find that fitting only the twist-six parameter is actually
enough to reach agreement with the experimental data.
This procedure is augmented by a more realistic
description of the hadronic content of the quasireal, i.e., the physical,
photon in terms of a spectral density that uses a Breit-Wigner (BW)
form to include the resonances of the $\rho$- and $\omega$-mesons.
The results of the fit are combined with the latest lattice constraints
from \cite{Bali:2019dqc} at the NNLO (two-loop) and N$^3$LO (three-loop) level
in conjunction with further constraints provided by
QCD sum rules with nonlocal condensates \cite{Bakulev:2001pa}, the aim being
to determine in this $Q^2$ regime appropriate values of the conformal coefficients
$b_2$ and $b_4$ of the twist-two pion distribution amplitude (DA).

The rest of the paper is organized as follows.
In Sec.\ \ref{sec:theor-basis} we present the new theoretical scheme
to calculate the pion-photon transition form factor within QCD.
This section encompasses the perturbative ingredients pertaining to
factorization and focuses on the implementation of the RG summation in
connection with a dispersion relation.
Sec.\ \ref{sec:mesonic-FAPT} discusses the TFF within the LCSR approach
in combination with ERBL summation, emphasizing the role of the hadronic
photon content of the LCSR.
The subsequent Sec.\ \ref{sec:data-fit1} is devoted to the processing of
the experimental data in the BESIII range, from 0.3 to 3.1~GeV$^2$,
in order to extract best-fit values of the nonperturbative scale parameters
$\delta^2_\text{tw-4}$, $\delta^2_\text{tw-6}$,
and the Gegenbauer coefficients $b_2$ and $b_4$.
A table with the BESIII data extracted from Fig.\ 3 in \cite{Redmer:2018uew}
using the tool PlotDigitizer~\cite{Rohatgi2020} is included.
The TFF predictions obtained with the new FAPT/LCSR scheme
are shown in comparison with a collection of data in a wider
momentum region up to $Q^2 < 5.5$~GeV$^2$ in Sec.\ \ref{sec:predictions}
making it apparent that our approach works well even above
the low-$Q^2$ range used in the fit.
Our conclusions are given in Sec.\ \ref{sec:concl}.
Some important calculational details are collected in four appendices.

\section{Theoretical basis of the $\pi-\gamma$ transition form factor}
\label{sec:theor-basis}
The pion-photon transition form factor
$F^{\gamma^*\gamma^*\pi^0}$ for two highly
virtual photons entering the reaction
$\gamma^*(-Q^2)\gamma^*(-q^2) \to \pi^0$
with virtualities
$Q^2, q^2 \gg m^2_\rho$
can be written by virtue of factorization as follows
\begin{subequations}
 \begin{eqnarray}
F^{\gamma^*\gamma^*\pi^0}\!(Q^2,q^2,\mu^2) \!\!&\sim&\!\!
T^{(2)}\!(Q^2,q^2,\mu^2;x)
\!\underset{x}\otimes\!
  \varphi_{\pi}^{(2)}\!(x,\mu^2)
\label{eq:TFFtw2}
~~~~~~
\\ &&  \hspace{-20mm}
+ T^{(4)}(Q^2,q^2,\mu^2;x)\underset{x}
\otimes
  \varphi_{\pi}^{(4)}(x,\mu^2)
  + \text{h.t.}\,,
\label{eq:TFFtw4-6}
\end{eqnarray}
 \end{subequations}
where h.t. abbreviates higher twist.
Here
$T^{(m)}$, related to the process
$\gamma^*\gamma^* \to q(G_{\mu\nu})\bar{q}$,
are perturbatively calculable hard-scattering parton amplitudes
entering convolutions with pion DAs
$\varphi_{\pi}^{(m)}$ of nonperturbative nature, where
$\underset{x}\otimes \equiv \int_0^1 dx$
and the superscript $(m)$ denotes the twist level of expansion.
To avoid unnecessary complications, the factorization (label F)
and renormalization (label R) scales have been set equal to each other
$\mu_\text{F}=\mu_\text{R}=\mu$ (default scale setting).
To perform the summation over the infinite series of the logarithmic
corrections, related to the renormalization of the coupling
$a_s=\alpha_s(\mu^2)/4\pi$
and the renormalization of the pion DA of leading-twist two
$\varphi_\pi^{(2)}(x,\mu^2)$,
we define a new running coupling
$\bar{a}_s(q^2\bar{y}+Q^2 y)\equiv \bar{a}_s(y)$
which also enters the ERBL exponent
\cite{Ayala:2018ifo}.
The ERBL exponent incorporates all evolution kernels $a^{k+1}_s V_k$,
whereas the partonic subprocesses, encoded in the coefficient functions
$a^k_s\mathcal{T}^{(k)}$,
are taken into account in terms of the leading-twist amplitude $T^{(2)}$.

\subsection{Main perturbative ingredients using RG summation}
\label{subsec:pert-RG-sum}
In order to carry out the RG summation, it is useful to expand
$\varphi_{\pi}^{(2)}(x,\mu^2)$, as well as the corresponding
contribution to the TFF in (\ref{eq:TFFtw2}), over the
conformal basis of the Gegenbauer harmonics
$\{\psi_n(x)=6x\bar{x}C^{3/2}_n(x-\bar{x}) \}$,
\begin{subequations}
\ba
  \varphi_{\pi}^{(2)}(x,\mu^{2})&=& \psi_{0}(x)
  + \sum_{n=2,4, \ldots}^{\infty} b_{n}(\mu^{2}) \psi_{n}(x),
\la{eq:gegen-exp}\\
  F^\text{(tw=2)}(Q^2,q^2)
&=& F_{0}^\text{(tw=2)}(Q^2,q^2)
\la{eq:gegen-FF}\\\nn &&
  + \sum_{n=2,4, \ldots}^{\infty} b_{n}(\mu^{2}) F_{n}^\text{(tw=2)}(Q^2,q^2) \, .
\ea
\end{subequations}
The partial form-factor contributions
$F_{n}^\text{(tw=2)}$ in the $\{\psi_{n} \}$ basis
in terms of the evolution exponential are given by
\begin{eqnarray}\nn
F_{n}^\text{(tw=2)}(Q^2,q^2) &=& N_\text{T} T_0(y)
\\\nn  && \hspace{-20mm}
	\underset{y}{\otimes}
	\left\{
		\vphantom{\int_{a_s}^{\bar{a}_s(y)}}
		\!\left[\1
            +\bar{a}_s(y)\mathcal{T}^{(1)}(y,x)
            +\bar{a}_s^2(y)\mathcal{T}^{(2)}(y,x)
            + \ldots
		\right]
	\right. ~~~~
\\\label{eq:Tfin} \!\!\!\! &&  \left. \hspace{-20mm}
    \underset{x}{\otimes}
    \exp\left[-\int_{a_s}^{\bar{a}_s(y)}\!\!
		       \frac{V(\alpha;x,z)}{\beta(\alpha)} d\alpha
	    \right]
\right\}
                   \underset{z}\otimes  \psi_n(z) \, .
\end{eqnarray}
Evaluating this expression at the one-loop level, its right hand side (RHS)
reduces to
\begin{eqnarray}\nn
&&F_{n}^\text{(tw=2)}(Q^2,q^2)\stackrel{\text{1-loop}}{\longrightarrow} F_{(1l)n}^\text{(tw=2)}
=
  N_\text{T}T_0(y)
\\ && \hspace{5mm}\label{eq:T1d}
  \underset{y}{\otimes}
  \left[\1+ \bar{a}_s(y)\mathcal{T}^{(1)}(y,x)\right]
   \left(\frac{\bar{a}_s(y)}{a_s(\mu^2)} \right)^{\nu_n}
   \underset{x}\otimes
   \psi_n(x) \, ,~~~
\end{eqnarray}
where $\mathcal{T}^{(1)}$ is the next-to-leading-order (NLO) coefficient
function and $T_0(y)$ is the Born term of the perturbative expansion
of $T^{(2)}$.
The other quantities entering (\ref{eq:T1d}) are the following
\begin{eqnarray} 
	\label{eq:various}
&&
	T_0(y) \equiv T_0(Q^2,q^2;y)
	=
	\frac{1}{q^2\bar{y}+Q^2 y}\, ,~
\\  &&
   	V(a_s;y,z) \to a_s V_0(y,z)\,,~
   	\beta(\alpha) \to a_s^2\beta_0 \,,~
\hspace{10mm}
\\\nn  &&
	\1 = \delta(x-y)\, ,~
	N_\text{T}=\sqrt{2}f_\pi/3\,,~
\\\nn &&  
	V_0(y,z)\otimes\psi_n(z)=-\frac{1}{2}\gamma_0(n)\psi_n(y)\,,
~~~~
\end{eqnarray}
where $V_0(y,z)$ is defined in Eq.~(\ref{eq:V}) and $a_s\gamma_{0}(n)$ denotes the one-loop anomalous dimension of
the corresponding composite operator of leading twist
with $\displaystyle\nu_n=\frac{1}{2}\frac{\gamma_0(n)}{\beta_0}$.
The next-to-next-to-leading-order (NNLO) expression for
$F_{(2)n}^\text{(tw=2)}$, analogous to Eq.\ (\ref{eq:T1d}),
is worked out in Appendix \ref{App:C}.

One notes that expression (\ref{eq:T1d}) does not contain the simple
product of the coupling
$\bar{a}_s^\nu(y) \equiv \bar{a}^\nu_s(q^2\bar{y}+Q^2 y)$
and the coefficient function
$\mathcal{T}(y,x)$, as usual, but their convolution.
For small values of $q^2$, this convolution has
for any $Q^2$ only a formal, not a physical meaning.
This becomes obvious from $T_0(Q^2,q^2;y)$, whose scale argument
$q^2\bar{y}+Q^2y$ approaches small values for $y\to 0$, even if
$Q^2$ is large, so that the perturbative expansion becomes unprotected.
This deficit is avoided, when a dispersion relation is involved.
As we show next, in this case, an equation like Eq.\ (\ref{eq:T1d})
can still be safely used in the TFF calculation
even for small $Q^2$ values.

\subsection{RG technique in connection with a dispersion relation}
\label{subsec:rg+disp-rel}
As we now demonstrate, summing over all radiative corrections in
Eq.\ (\ref{eq:T1d}), entails a new contribution to the imaginary part of
$F_{n}^\text{(tw=2)}(Q^2,-\sigma)$
and for the same reason also to the spectral density,
where $-\sigma$ is dual to $q^2$ \cite{Ayala:2018ifo}.
This marks an important difference to the standard version of the LCSRs
\cite{Khodjamirian:1997tk,Mikhailov:2009kf,Agaev:2010aq,Mikhailov:2016klg}.
To be specific, the imaginary part of the Born contribution is induced by
the singularity of $ T_0(Q^2,-\sigma;y)$ multiplied by a power of logarithms.
By contrast, the RG resummed radiative corrections lead to a term in the
spectral density that originates from the
$\Im\left(\bar{a}^\nu_s(-\sigma\bar{y}+Q^2y)/\pi\right)$ contribution.
We consider bellow the implementation of the RG summation in two steps,
starting with the same dispersion relation used in the LCSRs but
temporarily ignoring the hadronic content of the quasireal photon.
This will be taken into account in a subsequent step.

To start with, we go back to Eq.\ (\ref{eq:T1d}) and express $T_0(y)$ in
the form of a dispersion relation with respect to the variable
$q^2\rightarrow -\sigma$.
However, in contrast to the analogous result in \cite{Ayala:2018ifo}, we start
integrating at $m^2 \geqslant 0$, considering it as the threshold of
particle production.
This way, we obtain
\begin{widetext}	
\begin{eqnarray} \label{eq:1}
&& T_0(Q^2,q^2;y)\left(\bar{a}^{\nu_n}_s(y)\right) \otimes \psi_n(y)
	\stackrel{q^2\to -\sigma}{\longrightarrow}
	\frac{1}\pi \int_{m^2}^\infty d\sigma
   	\frac{\Im\big[T_0(Q^2,-\sigma;y)\bar{a}^{\nu_n}_s(-\sigma\bar{y}+Q^2y)\big]}{\sigma+q^2}
	\otimes \psi_n(y) =  I_n(Q^2,q^2)
\\ \nn
&&   =
	\frac{1}{\pi} \int_{m^2}^\infty \!\frac{d\sigma}{\sigma+q^2}
	\left\{
  		\Re[T_0(Q^2,-\sigma;y)]\Im[\bar{a}^{\nu_n}_s(-\sigma\bar{y}+Q^2y)]
	+ \Im[T_0(Q^2,-\sigma;y)]\Re[\bar{a}^{\nu_n}_s(-\sigma\bar{y}+Q^2y)]
	\right\}
	\otimes \psi_n(y)
\\ \label{2}
&& =\frac{1}{\pi}\int_{m^2}^\infty \!d\sigma
	\frac{\Re[T_0(Q^2,-\sigma;y)]\Im[\bar{a}^{\nu_n}_s(-\sigma\bar{y}+Q^2y)]}{\sigma+q^2}
	\otimes \psi_n(y)+\bm{0}\otimes \psi_n(y)\, .
\end{eqnarray}
The finite low integration limit modifies the result of the LCSR even at the
level of the Born term.
From a phenomenological point of view, $m^2$ can be assumed to be
$m^2=(2m_\pi)^2 \approx 0.078$ GeV$^2$, relating it to the pion pole,
or it can be considered as a tunable parameter.
\end{widetext}

Decomposing in Eq.\ (\ref{2}) the numerator
$T_0(Q^2,-\sigma;y)\sim 1/(-\sigma \bar{y}+Q^2y)$
and the denominator $\sigma+q^2$, while replacing the variables
$\sigma \to s= -(- \sigma\bar{y}+Q^2y) \geq 0 $,
one derives the integral
\begin{subequations}
\begin{eqnarray}
&& I_n(Q^2,q^2) = -\int_{m(y)}^\infty ds \frac{\rho_{\nu_n}(s)}{s(s+Q(y))} \otimes \psi_n(y) \,,
~~~
\end{eqnarray}
where
\begin{eqnarray}\nn
	&&\rho_\nu(s)=\Im[\bar{a}^\nu_s(-s-i\varepsilon)]/\pi\,,
	Q(y)\equiv q^2\bar{y}+Q^2y\,, ~~~\\
	&& 
	m(y)=m^2\bar{y}-Q^2y.
\end{eqnarray}
\end{subequations}
Keeping $m(y)>0$, we obtain
\begin{widetext}
\begin{eqnarray}
  I_n(Q^2,q^2)
	&=& -\left[\theta(m(y)> 0)\!\int_{m(y)}^\infty ds \frac{\rho_{\nu_n}(s)}{s(s+Q(y))}+
             \theta(m(y)\leqslant 0)\!\int_{0}^\infty\! ds \frac{\rho_{\nu_n}(s)}{s(s+Q(y))}\right]\otimes \psi_n(y) \nonumber\\
	&=&-\Big[\theta(m(y)> 0) J_{\nu_n}(m(y),Q(y)) +\theta(m(y)\leqslant 0) J_{\nu_n}(0,Q(y))\Big]\otimes \psi_n(y) \, ,
\label{eq:Idecompose}
\end{eqnarray}
\end{widetext}
where the second term corresponds to $m(y)\leqslant 0$ and the integral
starts at $s=0$.
The new terms $J_{\nu_n}(m(y),Q(y))$ can be recast in the form
\begin{subequations}
\label{4}
\begin{eqnarray}
\label{4a}
	-J_{\nu}(y,x)\!\!&=&\!\!- \int_{y}^\infty ds \frac{\rho_\nu(s)}{s(s+x)}
\\ \nn
	&=& \frac{1}{x}\left[\I_{\nu}(y,x)- \M_\nu(y)\right] \,,
\\ 
	\I_{\nu}(y,x)\!\!&\stackrel{\rm def}{=}&\!\! \int_{y}^\infty \frac{d\sigma}{\sigma+x} \rho_{\nu}^{(l)}(\sigma)\,,
\\ \nn
	\I_{\nu}(y \to 0,x)\!\!&=&\!\!\A_{\nu}(x),~\I_{\nu}(y,x \to 0)= \M_{\nu}(y),
\\ \label{4d}
	\I_{1}(y \to 0,x \to 0) &=& \A_{1}(0)=\M_{1}(0)\,,
\end{eqnarray}
\end{subequations}
where
$\I_{\nu}$
is a new coupling with $l$-loop content, introduced in
\cite{Ayala:2018ifo},
and
$\A_{\nu}$ \cite{Bakulev:2005gw}, $\M_{\nu}$ \cite{Bakulev:2006ex}
are the standard FAPT couplings in the spacelike and timelike regions,
respectively.
Use of (\ref{4a}) in (\ref{eq:Idecompose}) enables us to derive the
important expression
\begin{eqnarray}\nn 
&& \hspace{-7mm}
	I_n(Q^2,q^2) = T_0(Q^2,q^2;y)
\\\nn && \hspace{-5mm}
\times\Big\{
	\left[\I_{\nu}(m(y),Q(y))-\M_\nu(m(y))
	\right]
	\theta\left(m(y) > 0\right)
\\ && \hspace{0mm}
	+\left[\A_{\nu}(Q(y))-\M_\nu( 0)\right]\theta\left(m(y)\leqslant 0\right)
\Big\}\otimes \psi_n(y)\,, ~
\end{eqnarray}
in which the former couplings appear as limiting cases of $\I_{\nu}$,
cf.\ (\ref{4d}), while
\begin{eqnarray}\nn
\I_{\nu}(y,x)
	&=&\int_{y}^\infty \frac{ds}{s+x} \rho_{\nu}(s)
	=\A_{\nu}(x) -\!\!  \int_{0}^y \frac{ds}{s+x} \rho_{\nu}(s)
\\
	&=& \M_{\nu}(y) - x\int_{y}^\infty \frac{ds}{s(s+x)} \rho_{\nu}(s)\,
\end{eqnarray}
represents a generalized two-parameter coupling within FAPT
\cite{Ayala:2018ifo}.

Equipped with these results, we now consider the spectral density,
starting with the expression
\begin{eqnarray} \nn  
\rho_{\nu}^{(l)}(\sigma)
=
  \frac{1}{\pi}\,
  \textbf{Im}\,\big[a^{\nu}_{(l)}(-\sigma)\big]
= \frac{1}{\pi}\,\frac{\sin[\nu~
  \varphi_{(l)}(\sigma)]}{\left(R_{(l)}(\sigma)\right)^{\nu}}
\\
\stackrel{\text{1-loop}}{\longrightarrow} \frac{1}{\pi}\,
   \frac{\sin\left[\nu~\arccos\left(L_{\sigma}/\sqrt{L^2_\sigma+\pi^2}\right)\right] }{\beta_0^\nu~\left[\pi^2+L^2_\sigma\right]^{\nu/2}}
\end{eqnarray}
obtained in FAPT, where
$L_\sigma\equiv\ln(\sigma/\Lambda^2_{\rm QCD})$
and both the radial part $R_{(l)}$ and the phase $\varphi_{(l)}$
have a $l$-loop content, see \cite{Bakulev:2006ex}.
For our considerations below, it is useful to introduce a new effective coupling
$\bm{\mathbb{A}_\nu}$ by means of the parameter
$y_m=m^2/(m^2 +Q^2)$ to get
\begin{eqnarray}\nn
&&\mathbb{A}_\nu(m^2,y)
 =
 \theta\left(y\geqslant y_m\right) \left[\A_{\nu}(Q(y))-\M_\nu(0)\right]
\\\label{eq:eff-coupl} && \hspace{10mm}
+~\theta\left(y < y_m\right)
 \left[\I_{\nu}(m(y),Q(y))-\M_\nu(m(y))\right] \, . ~~~~~~
\end{eqnarray}
The coupling $\mathbb{A}_\nu(m^2,y)$ is a continuous function with respect to $y$ according to
(\ref{4d}).
In the applications to follow, we use in (\ref{eq:eff-coupl}) the zero-threshold
approximation $m^2\rightarrow 0$ so that
\be
 \mathbb{A}_\nu(m^2 \to 0,y) \to \mathbb{A}_\nu(0,y)=\A_{\nu}(Q(y))-\M_\nu( 0)\, ,
\ee
where the second term $\M_\nu( 0)=\A_\nu( 0)$ on the RHS demands some
care \cite{Ayala:2018ifo}, see
Sec.\ \ref{subsec:TFF-FAPT}.

\subsection{Pion-photon TFF in FAPT}
\label{subsec:TFF-FAPT}
Using Eq.\ (\ref{eq:eff-coupl}) in the limits
$q^2 \to 0$, $Q(y) \to yQ^2$, and $m^2 \geqslant 0$,
we derive for the TFF at one loop, cf.\ Eq.\ (\ref{eq:T1d}),
the following expressions
\begin{widetext}
\begin{subequations} \label{eq:TFFq0}
\begin{eqnarray} 
&&  \nu(n=0)=0; ~
Q^2 F^\text{(tw=2)}_{\text{FAPT},0}(Q^2)
		\equiv	F_0(Q^2;m^2)=
		N_\text{T}\left\{\int^1_{y_m}  \frac{\psi_0(x)}{x}~dx
		+ \left(\frac{\mathbb{A}_{1}(m^2,y)}{y}\right)
		\underset{y}{\otimes}
		\mathcal{T}^{(1)}(y,x)
		\underset{x}{\otimes}
		\psi_0(x)\right\} \, ,~~~~~
\\
&& \nu(n\neq0)\neq0; ~
Q^2 F^\text{(tw=2)}_{\text{FAPT},n\neq0}(Q^2)
	\equiv F_{n}(Q^2;m^2) =
\nonumber \\ && \hspace{35mm}
		\frac{N_\text{T}}{a_s^{\nu_n}(\mu^2)}
	\left\{
		\left(\frac{\mathbb{A}_{\nu_n}(m^2,y)}{y}\right)
		\underset{y}{\otimes}
		\psi_n(y)+\left(\frac{\mathbb{A}_{1+{\nu_n}}(m^2,y)}{y}\right)
		\underset{y}{\otimes}
		\mathcal{T}^{(1)}(y,x)
		\underset{x}{\otimes}
		\psi_n(x)
	\right\}.
\hspace{12mm}
	\end{eqnarray}
\end{subequations}
\end{widetext}
\mycomment{  
\begin{subequations}
\label{eq:TFFq0}
\begin{eqnarray}
\label{n0pTFFq0}
\!\!\!\!\!\!  \nu(n=0)=0;\!\!&\!\!&\!\! Q^2 F^\text{(tw=2)}_{\text{FAPT},0}
\equiv
  F_0(Q^2;m^2)
=
  N_\text{T}\left\{\int^1_{y_m}  \frac{\psi_0(x)}{x}~dx  \right.\nonumber \\
 &&\left. + \left(\frac{\mathbb{A}_{1}(m^2,y)}{y}\right)
\underset{y}{\otimes}
  \mathcal{T}^{(1)}(y,x)
\underset{x}{\otimes}
  \psi_0(x)\right\} \, ,\\
\!\!\!\!\!\!\nu(n\neq0)\neq0;\!\!&\!\!&\!\!  Q^2 F^\text{(tw=2)}_{\text{FAPT},n}
\equiv
  F_n(Q^2;m^2)
=
\nonumber \\
\!\!\!\!\!\!\!\!\!\!\!\!\!\!\frac{N_\text{T}}{a_s^{\nu_n}(\mu^2)}&\!\!&\!\!\!\!\!\!\!\!
  \left\{\left(\frac{\mathbb{A}_{\nu_n}(m^2,y)}{y}\right)
\underset{y}{\otimes}
  \psi_n(y)+\left(\frac{\mathbb{A}_{1+{\nu_n}}(m^2,y)}{y}\right)
\underset{y}{\otimes}
  \mathcal{T}^{(1)}(y,x)
\underset{x}{\otimes}
  \psi_n(x)\right\}.
\label{npTFFq0}
\end{eqnarray}
\end{subequations}
}
These equations can be reexpressed in the initial form of
Eq.\ (\ref{eq:T1d}) by performing a chain of substitutions that include
the zero-threshold $m^2 \to 0$ approximation and the replacements
$\left(y_m, \mathbb{A}_{\nu}(m^2,y)\right)
\stackrel{m=0}{\longrightarrow}\left( 0,\mathbb{A}_{\nu}(0,y) \right)$
$ \rightarrow (0,\bar{a}_s^\nu(y))$.
But, in contrast to Eq.\ (\ref{eq:T1d}), these expressions can be integrated
over $y$, because $\bm{\mathbb{A}_\nu}$ has \textit{no} Landau singularities.
This notwithstanding, singularities still appear at the origin $Q^2=0$ for
particular values of the index $\nu$, notably,  for $0< \nu < 1$.
In addition, for $\nu = 1$ at the upper bound of the interval $(0,1)$,
$\A^{}_{1}(0) = \M^{}_{1}(0) = 1/\beta_0 $ \cite{Shirkov:1997wi}
violates the asymptotic value of the TFF
$ Q^2F (Q^2 \rightarrow \infty) = \sqrt{2} f_\pi $, which is an exact result
of perturbative QCD in the asymptotic limit, see \cite{Lepage:1980fj}.
To fulfill it, we have to impose ``calibration conditions'' on the analytic
couplings and demand that \cite{Ayala:2018ifo}
\be \label{eq:calibration}
{\cal A }^{}_{\nu}(0)
={\mathfrak A}^{}_{\nu}(0)=0 ~~\text{for} ~~~0< \nu \leqslant 1 \, .
\ee
Let us mention that the models proposed in
\cite{Ayala:2016zrz,Ayala:2017tco}
comply with these conditions.

\subsection{About the role of NNLO corrections $O(\A_2)$ to the TFF}
\label{sec:NNLO}
Here we consider the NNLO$_\beta$ approximation of the partial form factors $F_n$
within FAPT pertaining to the standard RG expressions given in
Appendix \ref{App:C} in terms of
Eqs.\ (\ref{eq:Tevolve2}), (\ref{eq:apNNLOb0FOPT}).
The truncated series (\ref{eq:apNNLOb0FOPT}) of the powers $\bar{a}^{n}_s(y)$ in
the ERBL evolution factor can be easily mapped
into the same series by means of the replacement
$\bar{a}^{n}_s(y)\to \mathbb{A}_{n}(m^2,y)$
due to the linearity of the dispersion relation.
Applying the same calculational scheme as in
Sec.\ \ref{subsec:TFF-FAPT}
in the limits $q^2 \to 0$, $Q(y) \to yQ^2$,
we obtain from Eq.\ (\ref{eq:apNNLOb0FOPT}) the expression
\begin{widetext}
\begin{eqnarray}
\!\!\!\! Q^2 F_\text{FAPT;n}^\text{(tw=2)}(Q^2)
&\thickapprox&
  \frac{N_\text{T}}{\underline{\left[a_s(\mu^2)\right]^{\nu_n}} \left[1+c_1 a_s(\mu^2)\right]^{\omega_n } }
                                       \bigg\{
\underline{\frac{\mathbb{A}_{\nu_n }(m^2,x)}{x}
                                       +\left(\frac{\mathbb{A}_{1+\nu_n}(m^2,y)}{y}\right)\underset{y}\otimes\mathcal{T}^{(1)}(y,x)} \nonumber \\
&&+\omega_n c_1\left[\frac{\mathbb{A}_{1+\nu_n}(m^2,x)}{x}+ \frac{\mathbb{A}_{2+\nu_n}(m^2,x)}{x}\frac{c_1(\omega_n-1)}{2}
+ \left(\frac{\mathbb{A}_{2+\nu_n}(m^2,y)}{y}\right)\underset{y}\otimes \mathcal{T}^{(1)}(y,x)\right] \nonumber \\
&&+ \underline{\underline{
\left(\frac{\mathbb{A}_{2+\nu_n}(m^2,y)}{y}\right)\underset{y}\otimes
\mathcal{T}^{(2)}(y,x)}} \bigg\}
                           \underset{x}\otimes
                    \psi_n(x) \, ,
                                \label{eq:NNLO}
\end{eqnarray}
\end{widetext}
where the terms contributing in the leading logarithmic approximation (LLA),
cf.\ Eqs. (\ref{eq:TFFq0}), are underlined.
The couplings $a_s^\nu $ and $\mathbb{A}^{(l=2)}_{\nu }(m^2,x)$ should
be evaluated with a two-loop running, while $c_1=\beta_1/\beta_0$
and $\omega_n= [\gamma_1(n)\beta_0-\gamma_0(n)\beta_1]/[2\beta_0\beta_1]$.
The calculation of the FAPT couplings $\mathbb{A}^{(l=2)}_{2+\nu_n}$
with a two-loop running in Eq.\ (\ref{eq:NNLO}) is rather cumbersome.
Moreover, the couplings with the next higher index $2+\nu_n$ are
approximately an order of magnitude smaller than the couplings
$\mathbb{A}_{1+\nu_n}$ with a lower index.
We refrain from such a complicated and insignificant calculation here.
To estimate the effect of the
next-to-leading logarithmic approximation (NLLA),
it is sufficient to take into
account the contribution from the coefficient function
$\mathcal{T}^{(2)}(y,x)$ of the hard process in Eq.\ (\ref{eq:NNLO}),
denoted by the doubly underlined term in the third line
of Eq.\ (\ref{eq:NNLO}).
Only this term survives for the numerically important case of
the zero-harmonic, i.e., for $\omega_{n=0}=0$, while the terms
in the second line represent the effect of the two-loop ERBL-evolution.
For this reason, we use as a first estimate $c_1=0$.
To our knowledge, only the $\beta_0$ part of the two-loop ERBL
evolution is known \cite{Melic:2002ij}.
It is related to the contribution $\beta_0\mathcal{T}^{(2)}_{\beta_0} \to \mathcal{T}^{(2)}$
and enters the third line of Eq.\ (\ref{eq:NNLO}), see Appendix \ref{App:A}.
We can estimate the size of this effect by taking into
account the single contribution
\begin{eqnarray}
\label{eq:NNLOFAPT}
 \frac{N_\text{T}\beta_0}{\left[a_s(\mu^2)\right]^{\nu_n}}
 \!\left[\! \left(\frac{\mathbb{A}_{2+\nu_n}(m^2,y)}{y}\right)
 \!\underset{y}\otimes\!\mathcal{T}^{(2)}_{\beta_0}(y,x)\right]
                           \!\underset{x}\otimes\!
                    \psi_n(x)~~~
\end{eqnarray}
in addition to the LLA
in Eq.\ (\ref{eq:TFFq0}), keeping the evaluation of $\mathbb{A}_{2+\nu_n}(m^2,y)$
at the level of the one-loop running.

\section{Transition form factor within the LCSR employing ERBL summation}
\label{sec:mesonic-FAPT}
In the previous section we constructed a new perturbative expansion
that uses RG summation to include all radiative corrections to the TFF
while preserving its QCD asymptotics via calibration conditions.
In this section, we are going to implement this scheme to the LCSR formulation
of the TFF by means of the calibrated FAPT expansion.
Taking into account in the LCSR the hadronic content of the quasireal,
i.e., the physical, photon in terms of
the transition form factor $f_\rho F^{\rho \pi}$
in the spectral density \cite{Khodjamirian:1997tk,Mikhailov:2009kf}
\begin{equation}
  \rho^\text{ph}(s)
=
  \delta(s-m^2_\rho) \sqrt{2}f_\rho F^{\rho \pi}(Q^2)
  +\theta(s > s_0) N_{\text{T}}\rho(Q^2,s) \, ,
\label{eq:phys-spectral-dens}
\end{equation}
we get for $Q^2 F^{\gamma \pi}_\text{LCSR}$
\cite{Khodjamirian:1997tk,Mikhailov:2009kf,Agaev:2010aq},
the well-known expression
\begin{subequations}
\label{eq:BornLCSR}
\begin{eqnarray}\nn 
Q^2 F^{\gamma \pi}_\text{LCSR}\left(Q^2\right)
&=&
  N_{\text{T}}
	\left[\int_{x_s}^{1} \bar{\rho}(Q^2,\bar{x})
        \frac{dx}{x}
	\right.
\\  &&  \left. \hspace{-20mm}
 + \frac{Q^2}{m_{\rho}^2}
        \int_{0}^{x_{s}}
        \exp\left(
                  \frac{m_{\rho}^2-Q^2x/\bar{x}}{M^2}
            \right)
        \! \bar{\rho}(Q^2,\bar{x})
 	 \frac{dx}{\bar{x}}
 \right]
\\ 
&=&  N_{\text{T}}
  \left[H(Q^2) + \frac{Q^2}{m_{\rho}^2}V(Q^2,M^2)     \right],
\hspace{10mm}
\\  && \hspace{-28mm}
	\bar{\rho}(Q^2,x)=\varphi^{(2)}_{\pi}(x)
  	+ \bar{\rho}_\text{tw-4}(Q^{2}\!,x) +  \bar{\rho}_\text{tw-6}(Q^{2}\!,x)\, ,
\end{eqnarray}
where the integration variable in the spectral density
has been replaced by $s \to x=s/(Q^2+s)$ and $x_s=s_0/(Q^2+s_0)$.
Note that we use the $\delta$-resonance model
(\ref{eq:phys-spectral-dens}) only in order
to simplify the discussion, while the actual calculations
are performed by employing spectral densities
that include the resonances of the $\rho$- and $\omega$-mesons in the
form of a Breit-Wigner distribution,
see Appendix~\ref{App:D} and the discussion that follows.

The hard ($H$) and the soft ($V$) hadronic part of the TFF are
given, respectively, by
\begin{eqnarray}
H(Q^2)\!\!\!\!\!\!&& =\!\!
	\int_{x_s}^{1} \bar{\rho}(Q^2,\bar{x})
        \frac{dx}{x}\,,
\\
V(Q^2,M^2)\!\!\!\!\!\!&& =\!\!
	\int_{0}^{x_{s}}\!\!\!
        \exp\!\left(
                  \frac{m_{\rho}^2-Q^2x/\bar{x}}{M^2}
            \right)
        \! \bar{\rho}(Q^2,\bar{x})
  \frac{dx}{\bar{x}}\, .~~~~~
\end{eqnarray}
\end{subequations}
We use below the conformal expansion of the leading
twist-two part of $\bar{\rho}$,
expressing it in terms of the Gegenbauer harmonics to read
$
 \bar{\rho}(Q^2,x)
=
 \bar{\rho}_0(Q^2,x)+ \sum_{n=2,4,...} b_n(Q^2)\,\bar{\rho}_n(Q^2,x)$.
Moreover, we combine the twist-four and twist-six contributions
(see Appendix \ref{App:B})
with the $\psi_0$ component of the twist-two spectral density
into a single spectral density termed $\bar{\rho}_0$, i.e.,
\begin{subequations}
 \label{eq:rho}
\begin{eqnarray} 
\bar{\rho}_0(Q^2,x) &=& \psi_0(x) + \bar{\rho}_\text{tw-4}(Q^2,x)+\bar{\rho}_\text{tw-6}(Q^2,x)\,,
\hspace{7mm}
\\\label{eq:rho_0} 
\bar{\rho}_n(x)&=&\psi_n(x)\,.
\end{eqnarray}
\end{subequations}
The radiative contribution to the partial hard part $H_n$
contains the coupling
$\mathbb{A}_{\nu}(s_0;y)$, cf.\ (\ref{eq:eff-coupl}),
\begin{eqnarray}\label{eq:Heffect-m=0}
&&\mathbb{A}_{\nu}(s_0;y) =
	\theta \left(y\geqslant y_s \right)
	\left[\A_{\nu}(Q(y))-\M_\nu(0)\right]
\\ \nn  && \hspace{10mm}
 + \theta\left(y < y_s\right)\left[\I_{\nu}(s_0(y),Q(y))-\M_\nu(s_0(y))\right]\, ,
\quad{}
\end{eqnarray}
and was derived in
Sec.\,\ref{subsec:TFF-FAPT}.

On the other hand, the soft part $V_n$ contains the coupling
\begin{subequations}
\begin{eqnarray} \label{eq:Aeffect-m}
  \mathbb{A}_{\nu}(m^2;y)-\mathbb{A}_{\nu}(s_0;y)
=\theta(y_s >y )~\Delta_{\nu}(m^2,s_0;y)\,,~~~
\end{eqnarray}
where
\begin{widetext}
\begin{eqnarray} \nn 
  \Delta_{\nu}(m^2,s_0;y)
&=&\theta(y_s >y>y_m )\Big[\A_{\nu}\left(Q(y)\right)-\I_{\nu}\left(s_0(y),Q(y)\right)+\M_\nu\left(s_0(y)\right)-\M_\nu(0)\Big] \\
&~~~+\!&\!\theta(y_m >y )\Big[\I_{\nu}(m(y),Q(y))-\I_{\nu}(s_0(y),Q(y))+\M_\nu(s_0(y))-\M_\nu(m(y)) \Big].
\nonumber
\end{eqnarray}
\end{widetext}
Employing the zero-threshold approximation,
expression (\ref{eq:Aeffect-m}) 
reduces to
\begin{eqnarray}
\mathbb{A}_{\nu}(0;y)-\mathbb{A}_{\nu}(s_0;y) &=&
	\theta( y_s > y)~\Delta_{\nu}(s_0;y) \,,\nonumber
\\\nn
\Delta_{\nu}(s_0;y)&=&\A_{\nu}(Q(y))-\I_{\nu}(s_0(y),Q(y))
\\
&+&\M_\nu(s_0(y))-\M_\nu(0)\,
\label{eq:Aeffect-m=0}\, .
\end{eqnarray}
\end{subequations}
Combining the H-part, Eq.\ (\ref{eq:Heffect-m=0}), with the V-part,
Eq.\ (\ref{eq:Aeffect-m=0}), we obtain the total partial contribution
to the TFF within the FAPT/LCSR scheme
\begin{eqnarray} \nn
&& Q^2F^{\gamma\pi}_\text{LCSR/FAPT;n}\left(Q^2\right)=
	N_\text{T}
	\left[\phantom{\frac{Q^2}{m^2}}\hspace{-5mm} H_\text{FAPT;n}(Q^2)
	\right.
\\ \label{eq:HV-FAPT} && \hspace{20mm} \left.
	+\frac{Q^2}{m^2} k(M^2)\,V_\text{FAPT;n}(Q^2,M^2)\right]\, .~~~
\end{eqnarray}
To include the vector resonances into the spectral density entering the $V$-part,
we employ the more realistic Breit-Wigner formula
\cite{Khodjamirian:1997tk,Mikhailov:2009kf} rather than the simple
$\delta(\sigma -m^2_V)$ model.
This improved description of the soft part leads to the appearance of an
additional coefficient $k(M^2)$ in front of the term $V(Q^2,M^2)$
for the partial TFF $F^{\gamma\pi}_{\text{LCSR};n}\left(Q^2\right)$
in (\ref{eq:HV-FAPT}), see Appendix \ref{App:D} and \cite{Mikhailov:2009kf}.
Going one step further, we take into account the
$O(\mathbb{A}_2)$ contribution given in Eq.\ (\ref{eq:NNLOFAPT}) to derive
the following analytic expressions,
where the mentioned NLLA terms are
shown boldfaced in red color:
\begin{widetext}
\begin{subequations}
	\label{eq:finalFAPT/LCSR}
	\begin{eqnarray}
		\!\!\!\!\!\!\!\!\!\!\!\!\!\!\!\!\!\!\text{for~}n=0,~~~~~~~~~~~~
		H_\text{FAPT;0}(Q^2)&=&\int^{\bar{x}_s}_0\frac{dx}{\bar{x}}\bar{\rho}_0(Q^2,x)
		\nonumber \\
		&& +	\left(\frac{\mathbb{A}_{1}(s_0;x)}{x}
		\underset{x}{\otimes}
		\mathcal{T}^{(1)}(x,y)+\RedTn{\mathbf{ \frac{\mathbb{A}_{2}(s_0;x)}{x}
				\underset{x}{\otimes}
				\beta_0\mathcal{T}^{(2)}_\beta(x,y)}} \right)
		\underset{y}{\otimes}
		\psi_0(y) \,, \label{eq:finalFAPT/LCSR-H0}\\
		\!\!\!\!\!\!V_\text{FAPT;0}(Q^2,M^2)&=&
		\int^{1}_{\bar{x}_s}\frac{dx}{x}\exp\left(\frac{m_{\rho}^2}{M^2}- \frac{Q^2}{M^2}\frac{\bar{x}}{x}\right)
		\Bigg[\bar{\rho}_0(Q^2,x)
		\nonumber \\  \label{eq:finalFAPT/LCSRV0}
		&& + \left(\Delta_1(s_0,\bar{x})\mathcal{T}^{(1)}(\bar{x},y) +
		\RedTn{\mathbf{\Delta_2(s_0,\bar{x})\beta_0\mathcal{T}^{(2)}_\beta(\bar{x},y)}} \right)\underset{y}\otimes\psi_0(y)\Bigg],
		\label{eq:finalFAPT/LCSR-V0}
	\end{eqnarray}
	\begin{eqnarray}
		\!\!\!\!\!\!\!\!\!\!\!\!\!\!\!\!\!\!\text{for~}n>0,~~~~~
		H_\text{FAPT;n}(Q^2) &\!=\!&\!\!
		\frac{1}{a_s^{\nu_n}(\mu^2)}\left(
		\frac{\mathbb{A}_{\nu_n}(s_0;y)}{y} \right. \nonumber \\
		&\!\!&\!\!\left. + \frac{\mathbb{A}_{1+{\nu_n}}(s_0;x)}{x}
		\underset{x}{\otimes}
		\mathcal{T}^{(1)}(x,y)+ \RedTn{\mathbf{\frac{\mathbb{A}_{2+{\nu_n}}(s_0;x)}{x}
				\underset{x}{\otimes}\beta_0\mathcal{T}^{(2)}_\beta(x,y)}}
		\right)
		\underset{y}{\otimes}
		\psi_n(y),
		\label{eq:finalFAPT/LCSR-Hn}\\
		\!\!\!\!\!\!\!\!\!\!V_\text{FAPT;n}(Q^2,M^2)&\!=\!&\!\!\frac{1}{a_s^{\nu_n}(\mu^2)}
		\int_{\bar{x}_s}^1 \frac{dx}{x}
		\exp\left(\frac{m_{\rho}^2}{M^2}- \frac{Q^2}{M^2}\frac{\bar{x}}{x}\right)	
		\bigg[ \Delta_{\nu_n}(s_0,\bar{x})\psi_n(x)
		\nonumber \\  \label{eq:finalFAPT/LCSRVn}
		\!\!\!\!\!\!&& +\left(\Delta_{1+\nu_n}(s_0,\bar{x})
		\mathcal{T}^{(1)}(\bar{x},y) +\RedTn{\mathbf{\Delta_{2+\nu_n}(s_0,\bar{x})
				\beta_0\mathcal{T}^{(2)}_\beta(\bar{x},y)}}\right)\underset{y}\otimes\psi_n(y)\bigg]
		\, .
	\end{eqnarray}
\end{subequations}
\end{widetext}
Here the functions
$\Delta_{\nu_{n}}(s_0,\bar{x})$ and $\Delta_{1+\nu_{n}}(s_0,\bar{x})$,
defined in (\ref{eq:Aeffect-m=0}), represent effective couplings
entering the soft part with $\bar{x}_s\equiv 1-x_s=Q^2/(Q^2+s_0)$.
The hard partial contributions $H_\text{FAPT;n}$ in Eqs.\ (\ref{eq:finalFAPT/LCSR-H0}),
(\ref{eq:finalFAPT/LCSR-Hn}) coincide with the FAPT results given by
Eq.\ (\ref{eq:TFFq0}) after substituting $s_0$ by $m^2$
to get $s_0(y) \leftrightarrow m(y)$, $y_s \leftrightarrow y_m$.
Hence, the hard part of the process receives radiative corrections
driven by the same effective couplings, though
these corrections contribute at different thresholds $m^2$ and $s_0$.
On the other hand, the higher-twist contributions enter
(\ref{eq:finalFAPT/LCSR-H0}), (\ref{eq:finalFAPT/LCSR-V0}) by means of the term
$\bar{\rho}_0$ from Eq.\ (\ref{eq:rho_0}).
Let us emphasize that the evaluation of the perturbative contributions in
Eq.\ (\ref{eq:finalFAPT/LCSR}) at low momentum transfers is, in contrast to
the FOPT case, \emph{unrestricted}.

\section{Extraction of nonperturbative parameters from a data fit
at $Q^2 \leqslant 3.1$~GeV$^2$}
\label{sec:data-fit1}

\begin{figure*}[ht]
	\centerline{\includegraphics[width=0.47\textwidth]{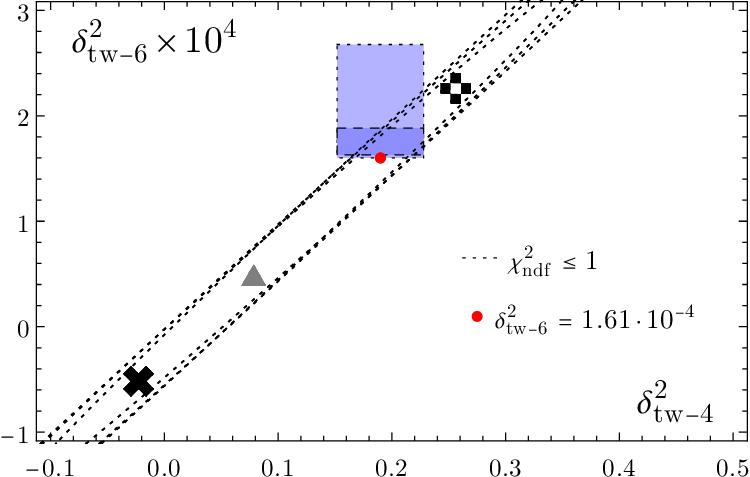}~
		\includegraphics[width=0.48\textwidth]{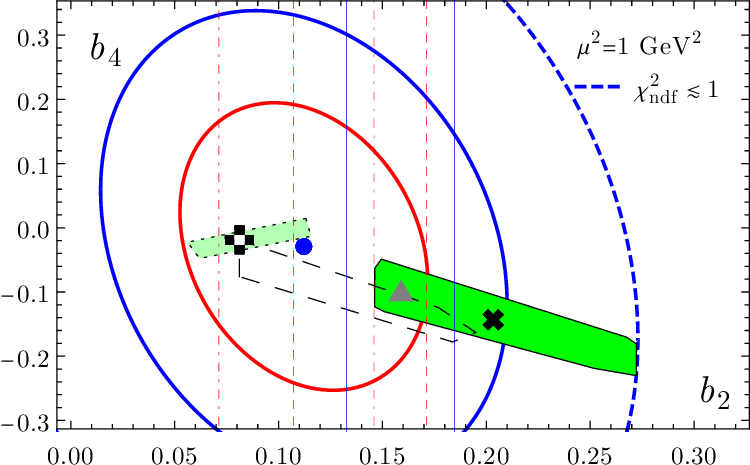}}
	\caption{\label{fig:fitTFF-NNLO} 
		Results of the two-step fitting procedure described in the text.
		The following notations are used:
		BMS DA---black cross \ding{54} \cite{Bakulev:2001pa};
		platykurtic DA---black/white cross \ding{60} \cite{Stefanis:2014nla,Stefanis:2015qha};
		DA shown as a grey triangle \GrayW{\ding{115}}
		selected from the BMS set of DAs to be inside the $1\sigma$ error ellipse (innermost red line)
		while fulfilling the N$^3$LO (vertical solid lines)
		lattice constraints on $b_2$ from lattice QCD \cite{Bali:2019dqc}.
		Left: first step of this procedure in which we determine the
		admissible regions of the higher-twist parameters $\delta_\text{tw-4}^2$ and
		$\delta_\text{tw-6}^2$.
		The larger rectangle denotes the range of values with
		$\delta_\text{tw-6}^2 =(1.84^{+0.84}_{-0.24})\,\times 10^{-4}\,\text{GeV}^6$,
		obtained in \cite{Cheng:2020vwr}, while the smaller rectangle
		corresponds to the estimate $\delta_\text{tw-6}^2 =(1.76\pm 0.13)\,\times 10^{-4}\,\text{GeV}^6$.
		This estimate and the red point $\delta_\text{tw-6}^2 = (1.61\pm 0.26)\times 10^{-4}$\,GeV$^6$ were
		obtained in this work.
		Right: results of the fitting procedure for the twist-two conformal coefficients
		$b_2,b_4$ with fixed higher-twist parameters.
		The two rectangles along the lower diagonal denote the range of $(b_2,b_4)$
		determined within the BMS approach \cite{Bakulev:2001pa} for two different values of
		$\lambda_q^2=0.4$~GeV$^2$ (larger shaded rectangle) and 0.45~GeV$^2$ (transparent rectangle), where
		the BMS DA \cite{Bakulev:2001pa} is represented by \ding{54}.
		The smaller shaded rectangle encloses the range of $(b_2,b_4)$ coefficients associated with
		DAs having a platykurtic profile \cite{Stefanis:2015qha}, like the model \ding{60} proposed
		in \cite{Stefanis:2014nla}.
		The dashed-dotted, dashed, and solid vertical lines show the lattice results
		for $b_2$ from \cite{Bali:2019dqc}
		for the NLO ($0.109(37)$), NNLO ($0.139(32)$),  and N$^3$LO ($0.159_{-0.027}^{+0.025}$), respectively.
	}
\end{figure*}

According to the exposition above, the domain of small $Q^2$ values
under single-tag conditions
becomes now accessible to a trustworthy perturbative description within
the FAPT/LCSR scheme using the TFF expression (\ref{eq:HV-FAPT}),
which involves resummed radiative corrections.
On the other hand, the higher-twist contributions can be safely included
within FOPT, see \cite{Mikhailov:2016klg,Stefanis:2020rnd}.
This allows for the first time a detailed and reliable
comparison with the recently released data with an unprecedented accuracy
below $Q^2=1.5$~GeV$^2$ of the BESIII experiment
\cite{Redmer:2018uew,Ablikim:2019hff}.
Because of the competitive accuracy up to 3.1~GeV$^2$ of these data,
it is possible to combine them with the measurements of previous single-tag
experiments, notably, CELLO \cite{Behrend:1990sr} and CLEO \cite{Gronberg:1997fj}
within the same range of momenta.
This way, we can perform a simultaneous best-fit analysis of these data sets
with the aim to determine the values of the involved nonperturbative
parameters in the calculation of the TFF.
These are the conformal coefficients $b_2$ and $b_4$ at twist-two
in Eq.\ (\ref{eq:gegen-exp}),
and the scale parameters for the twist-four, $\delta^2_\text{tw4}$, and the
twist-six, $\delta^2_\text{tw6}$, terms, given in Appendix \ref{App:B}.
At the normalization scale $\mu^2_0=1$~GeV$^2$, the mentioned parameters
assume values in the following ranges
\mycomment{
\begin{subequations}
\label{eq:npert-param}
\begin{eqnarray}
b_2(\mu_0^2)\! =\! [0.146, 0.272],\,\, b_4(\mu_0^2)\!=\! [-0.23, -0.049]
\label{eq:val-tw-2}
\end{eqnarray}	
for twist-two BMS DA domain~\cite{Bakulev:2001pa,Bakulev:2002uc,Bakulev:2004mc}
\begin{eqnarray}\label{eq:val-tw-4}
\delta_\text{tw-4}^2(\mu^2_0)=0.19\pm 0.04~\text{GeV}^2
\end{eqnarray}	
for twist-four~\cite{Bakulev:2002uc} and
\begin{eqnarray}
\begin{array}{l}
  \delta_\text{tw-6}^2(\mu^2_0)=(1.76\pm 0.13)\,\times 10^{-4}\,\text{GeV}^6 \text{(here)}\,,\label{eq:val-tw-6} \\
  \delta_\text{tw-6}^2(\mu^2_0)=\left(~1.84^{+0.84}_{-0.24}~\right)\,\,\times 10^{-4}\,\text{GeV}^6\, \text{\cite{Cheng:2020vwr}}\,,
\end{array}
\end{eqnarray}
for twist-six $\delta_\text{tw-6}^2(\mu^2_0)=\langle\sqrt{\alpha_s}\bar{q} q\rangle^2$.
\end{subequations}
}

\begin{widetext}
\begin{subequations}
	\label{eq:npert-param}
	\begin{eqnarray}
		\text{twist-two BMS DA domain~\cite{Bakulev:2001pa,Bakulev:2002uc,Bakulev:2004mc}}:&& %
		\Big\{b_2(\mu_0^2)\! =\! [0.146, 0.272],\,\, b_4(\mu_0^2)\!=\! [-0.23, -0.049] \Big\} \label{eq:val-tw-2} \\
		\text{twist-four~\cite{Bakulev:2002uc}}:&& \delta_\text{tw-4}^2(\mu^2_0)=0.19\pm 0.04~\text{GeV}^2
		\label{eq:val-tw-4} \\
		\text{twist-six}\,:&&\delta_\text{tw-6}^2(\mu^2_0)=
		\bigg\{
		\begin{array}{l}
			\langle\sqrt{\alpha_s}\bar{q} q\rangle^2=(1.76\pm 0.13)\,\times 10^{-4}\,\text{GeV}^6 \text{(here)}\label{eq:val-tw-6} \\
			\langle\sqrt{\alpha_s}\bar{q} q\rangle^2=\left(~1.84^{+0.84}_{-0.24}~\right)\,\,\times 10^{-4}\,\text{GeV}^6\, \text{\cite{Cheng:2020vwr}}\, .
		\end{array}
	\end{eqnarray}
\end{subequations}
\end{widetext}

\textbf{Fitting procedure---step 1.}
We perform a data fit that proceeds in two steps: First, we use Eqs.\
(\ref{eq:HV-FAPT}), (\ref{eq:finalFAPT/LCSR}) to determine
best-fit values of the higher-twist parameters
$\delta_\text{tw-4}^2,~\delta_\text{tw-6}^2$
keeping the twist-two nonperturbative coefficients $b_2$ and $b_4$ fixed,
see left panel of Fig.\ \ref{fig:fitTFF-NNLO}.
This gives rise to stretched out ellipses that degenerate into strips.
These strips for different DA models taken from the BMS-like
family, cf.\ (\ref{eq:val-tw-2})
overlap almost completely.
Such strips with $\chi^2_\text{ndf} \leqslant 1$ are displayed in the figure.
The uncertainties related to the higher-twist scales are shown graphically
in terms of shaded rectangles with respect to the central values of the
$\delta_\text{tw-4}^2$ and $\delta_\text{tw-6}^2$ scales in (\ref{eq:val-tw-6}).
The fitting is done by employing particular DAs with $b_2,b_4$ values
at 1~GeV obtained within QCD sum rules with nonlocal condensates:
\begin{enumerate}
\item
BMS model ($b_2=0.203$, $b_4=-0.143$) (\ding{54})---the center of the BMS domain \cite{Bakulev:2001pa},
\item
platykurtic (pk) $(b_2=0.0812, b_4=-0.0191)$ (\ding{60})
\cite{Stefanis:2014nla,Stefanis:2015qha}---see \cite{Stefanis:2020rnd} for further discussion,
\item
$(b_2=0.159, b_4=-0.098)$ \GrayW{\ding{115}}---crossing point of the N$^3$LO mean value from lattice \cite{Bali:2019dqc}
with the long axis of the BMS domain, see the right panel of Fig.\ \ref{fig:fitTFF-NNLO}
\end{enumerate}

These DAs can be used to fix the variations of the twist-two contributions and thus enable the
determination of the best-fit centers of the confidence ellipses for the scale coefficients
$\delta_\text{tw-4}^2,~\delta_\text{tw-6}^2$.
In fact, the main result of this fitting procedure is that all determined strips have a common long axis.
This implies that these parameters are strongly correlated and are aligned with this regression line.
On the other hand, this ascertained quasilinear dependence would entail an unpleasant overfitting of the
best-fit positions of $\delta_\text{tw-4}^2,~\delta_\text{tw-6}^2$ for the particular DAs.
Therefore, we proceed differently.
Using the mean value of $\delta_\text{tw-4}^2$ from  Eq.\ (\ref{eq:val-tw-4}), this  axis yields for the twist-six prefactor
the value $\delta_\text{tw-6}^2 = (1.61\pm 0.26)\times 10^{-4}$\,GeV$^6$,
where the error margin can be determined by imposing the condition $\chi^2_\text{ndf}\leqslant 1$.
This result bears no dependence on the choice of a particular model DA.
This $\delta_\text{tw-6}^2$ value---the red dot at the bottom of the rectangles in the figure---is in good
agreement with different independent estimates of $\delta_\text{tw-6}^2$ at its low limit.
This is outlined in Eq.\ (\ref{eq:val-tw-6}) and is discussed in more detail in
Appendix \ref{App:B}.
It corresponds to the dark violet rectangle in the left panel of Fig.\ \ref{fig:fitTFF-NNLO}.

\textbf{Fitting procedure---step 2.}
We can now use the best-fit values of the twist-four and twist-six parameters
($\delta_\text{tw-4}^2 =0.19$ GeV$^2$ and
$\delta_\text{tw-6}^2 =1.61\times 10^{-4}$\,GeV$^6$)
to derive the confidence regions of the twist-two expansion parameters
$b_2$ and $b_4$.
The results of this step of the fitting procedure are displayed in the
right panel of Fig.\ \ref{fig:fitTFF-NNLO}.
The best-fit point with $\chi^2_\text{ndf}=0.38$ is marked by the thick blue dot
($b^\text{b-f}_2=0.112$, $b^\text{b-f}_4=-0.029$),
whereas the $1\sigma$ error ellipse and the $2\sigma$ error ellipse are denoted by the innermost
red solid line and the blue solid line, respectively.
The outermost blue dashed ellipse corresponds to $\chi^2_\text{ndf}=1$.
In close correspondence to the left panel, we also show the positions of the
considered DA models at the normalization scale $\mu_0=1$~GeV.
The other ingredients of Fig.\ \ref{fig:fitTFF-NNLO}
(right panel) are the following:
\begin{itemize}
\item
Large shaded rectangle in green color---BMS domain
with the coefficients $b_2,b_4$ given above,
for the average virtuality of vacuum quarks $\lambda_q^2=0.40$~GeV$^2$ \cite{Bakulev:2001pa},
\item
Transparent rectangle bounded by a dashed line---domain of BMS DAs obtained for the larger virtuality $\lambda_q^2=0.45$~GeV$^2$.
\item
Small light-green rectangle---platykurtic range \cite{Stefanis:2015qha}
for the pk DA \ding{60} \cite{Stefanis:2014nla}.
\item
We also display for comparison, the most recent lattice constraints
on $b_2$ from \cite{Bali:2019dqc} using vertical lines.
These are obtained from left to right with NLO matching to \MSbar
scheme (dashed-dotted red lines), NNLO matching (dashed red lines),
and N$^3$LO, i.e., three-loop matching (solid blue lines).
This sequence of lines exhibits the progressive change of these constraints
as the loop order increases and the width of the corresponding strip decreases.
It is worth noting in this context that the various uncertainties of the lattice constraints
have been added in quadrature
which means that they are dominated by the largest
systematic error originating from the nonperturbative renormalization using the
regularization independent momentum subtraction (RI'/MOM) scheme
\cite{Bali:2019dqc,Martinelli_1995}.
\end{itemize} 
From this figure we can draw the following conclusions.

1.
The different sources of data
below $Q^2\leq 3.1$~GeV$^2$ give rise to a $2\sigma$ error ellipse (based on the NLLA)
which has a significant overlap with a large part of the BMS domain and also
the lattice constraints at the NNLO and N$^3$LO level,
restricting the values of $b_2$ to the range $[0.107, 0.184]$.

2.
A good portion of the BMS domain for $b_2^\text{BMS}<0.171$ lies
within the $1\sigma$ confidence ellipse of the data and also inside the N$^3$LO lattice strip.
This compatibility provides support to the BMS nonperturbative scheme and its ingredients.

3.
Imposing the most stringent combination of these constraints---$1\sigma$ ellipse
and N$^3$LO lattice range---one can determine a DA within the BMS domain defined by the crossing point
of the N$^3$LO lattice value $b_2=0.159$ (at $\mu^2_0=1$~GeV$^2$) with the long axis of the BMS rectangle
to obtain the value $b_4=-0.098$.
This uniquely defined DA with the parameters ($b_2=0.159,b_4=-0.098$)
provides a good compromise for the simultaneous fulfillment of
three distinct types of constraints originating from different sources.
It is denoted in Fig.\ \ref{fig:fitTFF-NNLO}
by \GrayW{\ding{115}} and is used in the following section to obtain predictions for the TFF.

4.
Note that an analogous crossing point of the NNLO middle point with the long BMS
line would fulfill the same requirements but would be outside the BMS rectangle.

5.
The platykurtic range lies entirely within the $1\sigma$
confidence ellipse and is close to the lattice NNLO strip.

\section{TFF predictions in the range $Q^2 \leqslant 5.5$~GeV$^2$ vs data}
\label{sec:predictions}
In this section, we present our TFF predictions obtained within the FAPT/LCSR
scheme developed in the previous sections.
We have two main objectives: To compare with various data up to an
intermediate momentum $Q^2=5.5$~GeV$^2$ and doing this to expose the improvements
relative to the FOPT/LCSR results.
Different approaches applicable to the calculation of the TFF in the low-$Q^2$ regime,
are mentioned in \cite{Stefanis:2020rnd}.
We include the full data sets of the BESIII \cite{Redmer:2018uew} and
CELLO \cite{Behrend:1990sr} Collaborations and also the measurements below 5.5~GeV$^2$
of the CLEO \cite{Gronberg:1997fj}, BaBar \cite{Aubert:2009mc}, and
Belle \cite{Uehara:2012ag} experiments.
The BESIII data with their errors have been extracted from
the graphics in Fig.\ 3 of \cite{Redmer:2018uew} using the tool
PlotDigitizer~\cite{Rohatgi2020} and are tabulated in Table
\ref{tab:BESIIIprelim}.
This restricted data selection is justified because our primary goal
is to show the utility of the summation technique in performing a LCSR calculation below/around 1~GeV$^2$.
At high $Q^2$, one can rely upon the FOPT/LCSR method, see \cite{Stefanis:2020rnd}
for such predictions and a complete list of the other data.
An alternative approach attempting to determine the higher moments
of the twist-two pion DA more reliably, was recently proposed in \cite{zhong2021improved}.
We note in similar context the Dyson-Schwinger-equations based approach recently
reviewed in \cite{Roberts:2021nhw} which uses a basis of Gegenbauer polynomials whose
degree is included in the optimization procedure to improve the convergence of the
polynomial expansion.

\begin{figure*}[t]
\centerline{\hspace{0mm}\includegraphics[width=0.7\textwidth]{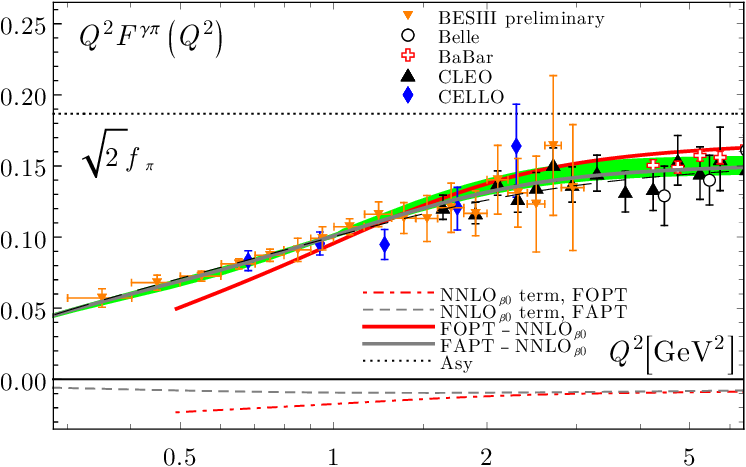}}
\caption{\label{fig:pionFF-strips}
Theoretical predictions for the scaled $\gamma^*\gamma\pi^0$ transition form factor
$Q^2F^{\gamma\pi}_\text{FAPT}(Q^2)$ [GeV] using different DAs discussed in the text in
comparison with various data up to $Q^2<5.5$~GeV$^2$ with labels as indicated in the figure.
The grey and the red solid lines were obtained with the DA denoted by \GrayW{\ding{115}}
using the FAPT/LCSR and FOPT/LCSR scheme, respectively.
The black dashed line represents the FAPT result obtained with the pk-DA \cite{Stefanis:2014nla},
while the green strip shows the theoretical uncertainties of the BMS DAs
calculated with
QCD sum rules with nonlocal condensates \cite{Bakulev:2001pa}.
The displayed FAPT/FOPT TFF results employ the best-fit nonperturbative higher-twist
parameters
$\delta_\text{tw-4}^2=0.19$\,GeV$^2$ and
$\delta_\text{tw-6}^2=1.61 \times 10^{-4}$\,GeV$^6$.\\
}
\end{figure*}

The TFF calculation is performed at the NNLO$_{\beta_0}$ for FOPT and in the NLLA for
FAPT (see Eqs.\ (\ref{eq:finalFAPT/LCSR})).
Using the expansion in Eq.\ (\ref{eq:gegen-FF}) and the partial
TFF terms $F^{\gamma\pi}_\text{LCSR/FAPT;n}$ from
Eqs.\ (\ref{eq:HV-FAPT}), (\ref{eq:finalFAPT/LCSR}), we obtain
predictions for $Q^2F^{\gamma\pi}_\text{FAPT}(Q^2)$ 
in terms of the conformal coefficients $\{1, b_2,  b_4,\ldots \}$
that can be used for any pion DA
(for a more detailed derivation at this level of accuracy, see
\cite{Mikhailov:2016klg,Stefanis:2020rnd} and references cited therein).
The results are shown graphically in Fig.\ \ref{fig:pionFF-strips}.

\begin{table*}[hbt]
	\begin{tabular}{lcccccccccccccccccc}\hline\hline
		$ Q^2$ [GeV$^2$]
		&0.351 & 0.45 & 0.551 & 0.652 & 0.751 & 0.851 & 0.951 & 1.075 & 1.226 &
		1.374 & 1.526 & 1.701 & 1.901 & 2.101 & 2.3 & 2.5 & 2.7 & 2.95
		\\\hline
		$\Delta Q^2$
		& 0.05 & 0.049 & 0.05 & 0.05 & 0.049 & 0.052 & 0.049 & 0.073 & 0.075 & 0.074
		& 0.075 & 0.101 & 0.1 & 0.099 & 0.101 & 0.099 & 0.1 & 0.15
		\\
		$Q^2 F(Q^2)$
		& 0.057 & 0.068 & 0.073 & 0.081 & 0.087 & 0.091 & 0.099 & 0.107 & 0.116 &
		0.113 & 0.113 & 0.121 & 0.117 & 0.14 & 0.131 & 0.123 & 0.164 & 0.135
		\\
		$Q^2 \Delta F(Q^2)$
		& 0.006 & 0.005 & 0.004 & 0.004 & 0.004 & 0.008 & 0.008 & 0.006 & 0.009 &
		0.011 & 0.016 & 0.017 & 0.016 & 0.024 & 0.024 & 0.034 & 0.049 & 0.044
		\\\hline\hline
	\end{tabular}
	\caption{ \label{tab:BESIIIprelim} \footnotesize
		Preliminary BESIII data on the scaled pion-photon TFF
		extracted from Fig.\ 3 in \cite{Redmer:2018uew}.
	}
\end{table*}

With reference to Fig.\ \ref{fig:fitTFF-NNLO}, we display
the TFF derived with the DAs from the BMS domain (large shaded rectangle) in the form of
a green strip with a variable width quantifying the variation of these
predictions entailed by the theoretical uncertainties of their key ingredients.
These are resolved at the bottom of the figure in order to give quantitative estimates
of their relevance (see the graphical explanations inside Fig.\ \ref{fig:pionFF-strips}).
Note that all displayed results are obtained by using in Eq.\ (\ref{eq:HV-FAPT})
the soft $V$-part given by Eqs.\ (\ref{eq:finalFAPT/LCSR-V0}), (\ref{eq:finalFAPT/LCSRVn})
and including the vector resonances $\rho$ and $\omega$ in the form of a Breit-Wigner
distribution, see Appendix \ref{App:D}.
This induces an additional factor that depends on the Borel parameter $M^2$,
taken to vary in the interval (0.75--1.1)~GeV$^2$ and depending on the momentum $Q^2$
as in \cite{Bakulev:2011rp, Bakulev:2012nh}.
The masses and widths of the vector resonances are
$m_\rho= 0.77$~GeV,
$\Gamma_\rho=0.1502$~GeV,
$m_\omega= 0.7826$~GeV,
$\Gamma_\omega=0.00844$~GeV.
The other LCSR parameters have been fixed in previous investigations to
the values \cite{Khodjamirian:1997tk,Bakulev:2002hk}
$f_\pi=0.132$~GeV,
$s_0 \approx 1.5~\text{GeV}^2$,
$\delta_\text{tw-4}^2(\mu^2_0)
= 0.95\, \lambda^2_q/2
=
 0.19~\text{GeV}^2$ \cite{Bakulev:2002uc} and are not varied here.
The twist-six scale
$
 \delta_\text{tw-6}^2(\mu^2_0)
=
 \langle\sqrt{\alpha_{s}}\bar{q} q\rangle^2(\mu^2_0)
=
 1.61 \times 10^{-4}\text{GeV}^6$
was determined in Sec.\ \ref{sec:data-fit1} by a fit to the experimental data
under the condition $\delta_\text{tw-4}^2=0.19\,\text{GeV}^2$, (see the previous section) and is approximately
equal to the lower bound of the estimates in (\ref{eq:val-tw-4}),~(\ref{eq:val-tw-6}).
Finally, the strong coupling $\alpha_{s}(\mu_{0}^{2})=0.48\pm 0.024$,
as well as the evolution of the DAs, are both taken in the two-loop
approximation, see Appendix A in \cite{Stefanis:2020rnd}.

The other displayed TFF predictions are the following.
The FAPT/LCSR TFF for the DA denoted by the symbol \GrayW{\ding{115}}, is shown
by the solid grey line,
while the analogous result for the FOPT/LCSR TFF is represented by the
solid red line.
For both curves the same values of the twist-four and twist-six parameters are used.
The displayed red curve serves only to demonstrate the tendency
of the FOPT/LCSR result to underestimate the data.
In fact, at 0.5~GeV$^2$ the calculated TFF is already outside the applicability domain
of this scheme.
In contrast, the FAPT/LCSR prediction, given by the light-grey line,
reproduces the data for momenta below $Q^2=5.5$~GeV$^2$ and down to values as low as $0.3$~GeV$^2$
with an accuracy of $\chi^2_\text{ndf}= 0.57$.
It is remarkable that the TFF calculated with the platykurtic DA (\ding{60})
\cite{Stefanis:2014nla}
(black dashed line)
turns out to be close to this line with $\chi^2_\text{ndf}= 0.77$.
This agrees with the results obtained recently within FOPT/LCSR in \cite{Stefanis:2020rnd}.
An important observation from
the curves shown at the bottom of
Fig.\ \ref{fig:pionFF-strips} is that above $Q^2>2$~GeV$^2$, the NNLO$_{\beta}$
parts of both LCSR schemes (FOPT--dashed-dotted line and
FAPT---dashed line) yield congruent results.
Below $Q^2\lesssim 1$~GeV$^2$, the RG summation of the
radiative corrections (dashed line at the bottom) in the FAPT/LCSR scheme
avoids the overestimation of the NNLO correction in the FOPT/LCSR scheme
(dashed red line at the bottom), clearly demonstrating its superiority.

\section{Conclusions}
\label{sec:concl}
In this work we developed and outlined a new theoretical scheme to
calculate the pion-photon transition form factor with single-tag kinematics
that involves RG summation of radiative corrections while avoiding Landau
singularities of the running strong coupling.
We showed that this scheme, termed FAPT/LCSR, is capable of providing
trustworthy results well below the typical hadronic scale of 1~GeV,
a regime not reliably accessible using FOPT/LCSR.
This allows the comparison of theoretical predictions with the recently
released preliminary data of the BESIII Collaboration \cite{Redmer:2018uew} which bear very small
errors just in this momentum region.

To include the hadronic content of the quasireal photon, we used
in the phenomenological part of the LCSR a Breit-Wigner distribution which provides a more
realistic representation than a simple $\delta$-function ansatz.
This admits the possibility of comparing more precisely
the obtained TFF predictions with those in the state-of-the-art analysis within
FOPT/LCSR in \cite{Stefanis:2020rnd}, which also employs the Breit-Wigner form.
This way, the effect of including the QCD radiative corrections by means of RG
summation has been properly determined.
Doing so, we were able to substantially exceed our
exploratory analysis in \cite{Ayala:2018ifo,Ayala:2019etj}
and promote our understanding of the TFF behavior at much lower momentum scales.
In the following we collect and discuss further the key results of our analysis.

(i)
We used the available experimental data in the
low-momentum domain up to $Q^2\leq 3.1$~GeV$^2$ in order to determine
best-fit values of the higher-twist parameters.

Especially the measurement of the BESIII experiment \cite{Redmer:2018uew}
provided data with an unprecedented accuracy below $Q^2=1.5$~GeV$^2$.
Using this data set in combination with previous data of the CELLO \cite{Behrend:1990sr} and CLEO \cite{Gronberg:1997fj}
Collaborations, we obtained a reliable estimate for the
twist-six contribution
$\delta_\text{tw-6}^2=1.61 \times 10^{-4}$\,GeV$^6$
using $\delta_\text{tw-4}^2=0.19$\,GeV$^2$ from ~\cite{Bakulev:2002uc}
and keeping the conformal coefficients
$b_2$ and $b_4$ within the BMS domain.

(ii)
In the second step, we used these parameters to extract the
most trustworthy regimes of the conformal coefficients ($b_2,b_4$) by applying additional constraints
from the data and the most recent lattice calculations of $b_2$.
To be precise, we determined the $1\sigma$ and $2\sigma$ error ellipses of the data and combined them with the
lattice constraints of \cite{Bali:2019dqc} at the NNLO, and N$^3$LO level.
Combing these constraints in the most stringent way, we found that the crossing point
of the middle value of the N$^3$LO lattice range of $b_2$ with
the long axis of the BMS domain \cite{Bakulev:2001pa}
of the ($b_2,b_4$) values, defines a DA,
marked by the symbol
\GrayW{\ding{115}}, that agrees with the employed data at the $1\sigma$ level.

(iii)
Employing this DA as nonperturbative input, we performed a twin-calculation of the TFF in
FAPT/LCSR and in FOPT/LCSR in order to quantify the advantage of including the radiative corrections
via RG summation.
The corresponding TFF curves are shown in Fig.\ \ref{fig:pionFF-strips} in terms of a grey and a red curve, respectively.
One appreciates that the FAPT result reproduces	the data in a momentum range starting below $1$~GeV$^2$
and extending up to 5.5~GeV$^2$ at the level of an overall accuracy of $\chi^2_\text{ndf}=0.57$.

(iv)
The FAPT/LCSR TFF result for the platykurtic pion DA \cite{Stefanis:2014nla},
shown as a black dashed line in Fig.\ \ref{fig:pionFF-strips}, follows closely the grey curve and the BMS
strip (in green color) of predictions though this DA has a unimodal profile in contrast
to the bimodal shapes of the BMS DAs.
This can be traced to the values of their inverse moments that almost coincide.
For the discussion of the properties of this DA, we refer to \cite{Stefanis:2020rnd}.

As a last remark, we mention that our exposed method may be useful in providing
insight into the hadronic light-by-light contribution of the g-2 of the muon, see
\cite{Masjuan:2012wy,Dorokhov:2014iva} and \cite{Aoyama:2020ynm}
for a recent review.
Moreover, a pion DA very close to \GrayW{\ding{115}} was used very
recently in \cite{leljak2021barbto} (see Table 1) to calculate the $\bar{B}\rightarrow \pi$
form factors and determine $|V_{ub}|$ in agreement with inclusive estimates.

\acknowledgments
S.~V.~M. acknowledges support from the Heisenberg--Landau Program 2020.
A.~V.~P. was supported by the Strategic Priority Research Program of the Chinese Academy of Sciences,
Grant No.\ XDB34000000 and the President's International Fellowship Initiative (PIFI Grant No.\ 2019PM0036).
The work of A.~V.~P. was carried out with the financial support from the Ministry of Science and
Higher Education of the Russian Federation (State task in the field of scientific activity,
scientific project No.\ 0852-2020-0032, (BAZ0110/20-3-07IF)).
A.~V.~P. thanks A.~Radzhabov, Yu.~Markov and A.~Kaloshin for fruitful discussions and the
warm hospitality at the Institute for System Dynamics and Control Theory of the Siberian Branch
of the Russian Academy of Sciences and the Theoretical Physics Department of the Irkutsk State University.

All authors contributed equally to this work.

\begin{appendix}
\appendix
\section{QCD perturbative expansion beyond leading order}
\label{App:A}
\textbf{NLO.}
The coefficient function of the partonic subprocess
$\mathcal{T}^{(1)}$ is
\begin{subequations}
\begin{eqnarray} 
  \frac{{\cal T}^{(1)}(x,y)}{C_{\rm F}}
&=&
  \left[-3 V^{b}  +  g \right]_+(x,y) - 3 \delta(x-y),
\\\nn 
  g_{+}(x,y)
&=&
  -2\left[
         \theta(y>x)\frac{\ln\left(1-x/y\right)}{y-x}
\right.\\  &&  \hspace{7mm} \left.
		+\theta(y<x)\frac{\ln(1-\bar{x}/\bar{y})}{x-y}
    \right]_{+},~~~
\end{eqnarray}
\end{subequations}
while the elements of the one-loop evolution kernel $V_0$ are
\begin{subequations}
\begin{eqnarray}\nn
 \frac{V_0(x,y)}{C_{\rm F}}  &=& V^{(0)}_+(x,y) =
  2\left[{\cal C}\theta(y>x)\frac{x}{y}
         \left(1+\frac{1}{y-x}\right)
         \right]_{+}
\\\label{eq:V} &\equiv&
	2 \left[V^{a}(x,y) +V^{b}(x,y)\right]_{+}\, ,
\end{eqnarray}
\begin{eqnarray}
\nn 
V^{a}(x,y) &=&
   {\cal C} \theta(y>x)\frac{x}{y},
\\
V^{b}(x,y) &=&
  {\cal C} \theta(y>x)\frac{x}{y}\left(\frac{1}{y-x}\right)\, ,
\end{eqnarray}
\end{subequations}
where the symbol $\mathcal{C}$ means
$ {\cal C}=\1+\left\{ x \to \bar{x}, y \to \bar{y}\right\} $.
The  key term of the  convolution ${\cal T}^{(1)}(x,y)\otimes\psi_n(y)$
that enters the harmonic expansion can be significantly simplified to get
\begin{widetext}
\begin{eqnarray}
\frac{1}{C_{\rm F}} {\cal T}^{(1)}(x,y)\underset{y}\otimes\psi_0(y) &=&
    \left[ -3+\frac{\pi^2}{3} - \ln^2\left(\frac{\bar{x}}{x}\right) \right] \psi_0(x) - 2~\psi_0(x),
\\  
\frac{1}{C_{\rm F}} {\cal T}^{(1)}(x,y)\underset{y}\otimes\psi_n(y)&=&
    \left[ -3\left(1+v^{b}(n)\right)+\frac{\pi^2}{3} - \ln^2\left(\frac{\bar{x}}{x}\right) \right] \psi_n(x)
    - 2   \sum^n_{l=0,2,\ldots}\!\!\!G_{nl}\psi_l(x) \, ,
\\ 
v^{b}(n)&=& 2\left(\psi(2)-\psi(2+n) \right);~v(n)=1/\left[(n+1)(n+2)\right]-1/2 + 2\left(\psi(2)-\psi(2+n) \right)\,,
\end{eqnarray}
\end{widetext}
see Appendix A in \cite{Mikhailov:2009kf}.
The quantities $v^{b}(n)$ and $v(n)=-\frac{1}{2 C_{\rm F}}\frac{1}2\gamma_0(n)$ are the eigenvalues of the
elements $V^b_+$ and $V^a_+ + V^b_+$ of the one-loop kernel in
Eq.\ (\ref{eq:V}), respectively.
$G_{nl}$ denotes the elements of a calculable triangular matrix
(omitted here)---see for details \cite{Mikhailov:2009kf} and corrections in \cite{Agaev:2010aq}.

The explicit expressions for the first coefficients of the expansion of the QCD $\beta$ function are
\begin{widetext}
\begin{eqnarray}
   &&\beta_0 = \frac{11}{3}\,C_\text{A} - \frac{4}{3}\,T_r N_f
    \, ,\qquad
\beta_1 = \frac{34}{3}\,C_{\text{A}}^{2}
        - \left(4C_\text{F}
        + \frac{20}{3}\,C_\text{A}\right)T_r N_f\, .
\end{eqnarray}
\end{widetext}

\textbf{NNLO.}
The $\beta_0$ part of the coefficient function ${\cal T}^{(2)}$, the $\beta_0{\cal T}^{(2)}_{\beta}$ term, reads
\begin{widetext}
\begin{eqnarray} 
 {\cal T}^{(2)}_{\beta}(x,y)
&=&
  C_{\rm F}\Bigg[\frac{29}{12} 2V^{a} + 2\dot{V}^{a}
        - \frac{209}{36} V^{(0)}  - \frac{7}{3} \dot{V}^{(0)}
        - \frac{1}{4} \ddot{V}^{(0)} + \frac{19}{6} g
        + \dot{g}
  \Bigg]_+\!\!(x, y)  - 6 C_{\rm F}  \delta(x-y)\, .
\end{eqnarray}
\end{widetext}

This expression was originally derived in
\cite{Melic:2002ij}, but its elements are presented here using a different
notation following \cite{Mikhailov:2009kf}, where also the omitted explicit expressions for the elements
$\dot{V}^{a}, \dot{V}^{(0)}, \ddot{V}^{(0)}$ and $\dot{g}$ can be found.

\section{Higher-twist contributions}
\label{App:B}
The explicit expressions for the twist-four and twist-six \cite{Agaev:2010aq} contributions are given by
\begin{eqnarray}
&&
	\bar{\rho}_\text{tw-4}(Q^2,x)=\frac{\delta_\text{tw-4}^2(Q^2)}{Q^2}x\frac{d}{dx}\varphi^{(4)}(x);
\\\nn
&&
	\varphi^{(4)}(x)=\frac{80}3 x^2(1-x)^2\, ,
\\&&
	\delta_\text{tw-4}^2(Q^2)
   		= \left[\frac{a_s(Q^2)}{a_s(\mu^2_0)}
     		\right]^\frac{\gamma_{T4}}{\beta_0}\delta_\text{tw-4}^2(\mu^2_0)\,,
\\\nn &&
	\delta_\text{tw-4}^2(\mu^2_0)= 0.95\, \lambda^2_q/2 = 0.19~\text{GeV}^2 ~\text{\cite{Bakulev:2002uc}}\,, \gamma_{T4} = 32/9\, ,
\end{eqnarray}
\begin{eqnarray}
\nn &&
	\bar{\rho}_\text{tw-6}(Q^{2}\!,x)  =
    8\pi \frac{C_F}{N_c}
    \frac{  \alpha_s\langle\bar{q} q\rangle^2 }{f_\pi^2}\frac{x}{Q^4}
    \left[
        \!-\!
        \left(\frac{1}{1-x}\right)_+
	\right.
\\ && \hspace{23mm}
	\left.
        \!+\!\left(2\delta(\bar{x})-4 x\right)\!+\!
        x\left(
         3+2\ln(x\bar{x})
               \right)
    \phantom{\left(\frac{1}{x}\right)_+}\hspace{-10mm}
    \right] \, ~~~~~
\end{eqnarray}
at $\mu^2_0=1$\,GeV$^2$.
\begin{subequations}
The NLO evolution of the quark condensate $\langle\bar{q} q\rangle(\mu^2)$ reads
\begin{eqnarray}\nn
\langle\bar{q} q\rangle(\mu^2)&=& \langle\bar{q} q\rangle(4\,\text{GeV}^2)
	\left[
            \frac{\bar{a}_{s}^{(2)}(4\,\text{GeV}^2)}{\bar{a}^{(2)}_s(\mu^2)}
    \right]^{\nu} \!
\\\label{eq:cond2-levo} && \hspace{8mm}
      \times\left[\frac{1+c_1\bar{a}_{s}^{(2)}(4\,\text{GeV}^2)}{1+c_1\bar{a}^{(2)}_s(\mu^2)}
      \right]^{\omega} ~~
\end{eqnarray}
 for $N_f=3$, $\Lambda^{(3)}_\text{MS}=392$ MeV with a two-loop running, where
\begin{eqnarray}
&&
	\nu= \gamma_0/\beta_0\,, ~\gamma_0=4\,, ~\beta_0=11-\frac{2}{3}N_f\,,
\\ &&
	\beta_1=102-\frac{38}{3}N_f\,, ~\omega=[\gamma_1\beta_0-\gamma_0\beta_1]/[\beta_0\beta_1]\,,
~~~~~~\\ &&
	\gamma_1= \frac{202}3 -\frac{20}9 N_f\,, ~c_1=\beta_1/\beta_0 \, .
\end{eqnarray}
\end{subequations}
The quark-condensate density is obtained
from the Gell-Mann-Oakes-Renner relation
\begin{eqnarray}
  \langle\bar{q} q\rangle(\mu^2)&=&- \frac{f_\pi^2 m^2_\pi}{2(m_u+ m_d)(\mu^2)} \, .
\end{eqnarray}
Taking into account the estimate
$(m_u+m_d)(\mu^2_L=4\text{GeV}^2)= 7.28(82)~\text{MeV}$
from lattice computations \cite{Tanabashi:2018oca} and performing a two-loop running
according to (\ref{eq:cond2-levo}),
we obtain for a ``quasi'' (one-loop) RG invariant quantity the new estimate
\begin{subequations}
\label{eq:q-cond-latt}
\begin{eqnarray}
  \langle \sqrt{\alpha_{s}}\bar{q} q\rangle^2(\mu^2_0)&=&(1.76\pm 0.13)\,\times 10^{-4}\,\text{GeV}^6\,,~~~~
  \\\nn
  \alpha_{s}(\mu^2_0)&=&0.48\pm 0.024 \, ,
\end{eqnarray}
by employing a NLO approximation.
A result for this quantity close to that was found in \cite{Cheng:2020vwr}
within a similar approach using the RunDec code, see Fig.\ \ref{fig:fitTFF-NNLO} (left panel):
\begin{eqnarray}
  \langle \sqrt{\alpha_{s}}\bar{q} q\rangle^2(\mu^2_0)&=&(1.84^{+0.84}_{-0.24})\,\times 10^{-4}\,\text{GeV}^6\,,~~~~
\\\nn
  \alpha_{s}(\mu^2_0)&=&0.486\pm 0.024\, .
\end{eqnarray}
\end{subequations}
Our best-fit estimate $\delta_\text{tw-6}^2 = \langle \sqrt{\alpha_{s}}\bar{q} q\rangle^2(\mu^2_0)= (1.61\pm 0.26)\times 10^{-4}$\,GeV$^6$
overlaps within errors with both values given above.

\section{ERBL summation at NNLO}
 \label{App:C}
The conformal symmetry of the ERBL equation at the two-loop level of evolution is broken.
This entails within the basis of Gegenbauer harmonics the appearance of a nondiagonal part
along with the dominating diagonal one \cite{Mikhailov:1984ii,Mikhailov:1985cm}.
We consider here the most significant \textit{diagonal part} at the two-loop level
of the evolution exponential in Eq.\ (\ref{eq:Tfin}), notably,
\begin{eqnarray}\nn 
&& \hspace{-17mm} \exp\left[-\int\limits_{a_{s}(\mu^2)}^{\bar{a}_{s}(y)}\!\!
                                                    \frac{V(\alpha;x,z)}{\beta(\alpha)} d\alpha
              \right]\underset{z}\otimes  \psi_n(z)
\\\nn \stackrel{2\,\text{loop}}{\longrightarrow} &&
    \exp
      \left[
            \int\limits_{a_{s}(\mu^2)}^{\bar{a}_{s}(y)}
            \frac{\gamma_n(a)}{\beta(a)}da
      \right]\psi_n(x)=
\\ &&
      \left[
            \frac{\bar{a}_{s}(y)}{a_s(\mu^2)}
      \right]^{\nu_n}
      \left[\frac{1+c_1\bar{a}_{s}(y)}{1+c_1a_s(\mu^2)}
      \right]^{\omega_n}\psi_n(x) \, ,
\end{eqnarray}
where
$a_s$ has a two-loop running
and  $c_1=\beta_1/\beta_0$, with $\beta_i$ being the expansion coefficients of the
QCD $\beta$-function.
The evolution exponent of the coupling is defined by
$\displaystyle\nu_n=\gamma_0(n)/2\beta_0$,
$
 \omega_n
=
 [\gamma_1(n)\beta_0-\gamma_0(n)\beta_1]/[2\beta_0\beta_1]
$.
The corresponding diagonal part of the partial form factors
$F_{n}^\text{(tw=2)}$ (in the $\{\psi_{n} \}$ basis) has the form
\begin{widetext}
\begin{eqnarray}\label{eq:Tevolve2}
\!\!\!\!  F_{n}^\text{(tw=2)}(Q^2,q^2)
=
  N_\text{T} T_0(y) \underset{y}{\otimes}
                                        \!\left[\1
                                                  +\bar{a}_s(y)\mathcal{T}^{(1)}(y,x)
                                                  +\bar{a}_s^2(y)\mathcal{T}^{(2)}(y,x)
                                          \right]
\left(\frac{\bar{a}_s(y)}{a_s(\mu^2)} \right)^{\nu_n} \left[ \frac{1+c_1\bar{a}_s(y)}{1+c_1 a_s(\mu^2)}\right]^{\omega_n}
                           \underset{x}\otimes
                    \psi_n(x) \, .
\end{eqnarray}
\end{widetext}
Every harmonic $\psi_n$ generates under the two-loop evolution the contribution of off-diagonal
higher harmonics \cite{Mikhailov:1985cm}, but these are small compared to the diagonal ones.
Therefore, they are not considered here.
Moreover, we use the following approximation to (\ref{eq:Tevolve2})
\begin{widetext}
\begin{eqnarray}
\!\!\!\!  F_{n}^\text{(tw=2)}(Q^2,q^2)
&\thickapprox&
  \frac{N_\text{T}}{\underline{\left[a_s(\mu^2)\right]^{\nu_n}}\left[1+c_1 a_s(\mu^2)\right]^{\omega_n} } T_0(y) \underset{y}{\otimes}
                                        \!\bigg\{\underline{\left[\bar{a}_s(y)\right]^{\nu_n}+\left[\bar{a}_s(y)\right]^{1+\nu_n}\mathcal{T}^{(1)}(y,x)} \nonumber \\
&&
        +~\omega_n c_1\left[ \left[\bar{a}_s(y)\right]^{1+\nu_n}+\left[\bar{a}_s(y)\right]^{2+\nu_n}\frac{(\omega_n-1)}{2} c_1 + \left[\bar{a}_s(y)\right]^{2+\nu_n}\mathcal{T}^{(1)}(y,x) \right] \nonumber \\
&&      + \left[\bar{a}_s(y)\right]^{2+\nu_n}\mathcal{T}^{(2)}(y,x)\bigg\}
                           \underset{x}\otimes
                    \psi_n(x) \, ,
\label{eq:apNNLOb0FOPT}
\end{eqnarray}
\end{widetext}
where the last factor $\left[1+c_1 \bar{a}_s(y)\right]^{\omega_n}$ in (\ref{eq:Tevolve2}) has been expanded.
The additional new terms are presented in the second and the third line of Eq.\ (\ref{eq:apNNLOb0FOPT}),
while the terms with a ``NLO structure'', cf.\ Eq.\ (\ref{eq:T1d}), in the first line are underlined.

\section{Pion TFF in LCSR with a Breit-Wigner resonance model}
 \label{App:D}
We present the original expression \cite{Ayala:2018ifo}
as the sum of a hard term $H$ and a vector-resonance term $V$
\begin{eqnarray}\nn
 F^{\gamma\pi}_{\text{LCSR};n}\!\left(Q^2\right)
\!=\!
  \frac{N_\text{T}}{Q^2}\!\left[H_{n}(Q^2)\!+\!\frac{Q^2}{m^2_\rho} k(M^2)V_{n}(Q^2,M^2)\right],
  \!\!\!\!
\\\label{eq:AppB-HV}
\end{eqnarray}
where the coefficient $k$ has been introduced to modify the original expression \cite{Ayala:2018ifo}
by modeling the rho-meson resonance contribution in terms of a BW distribution instead of a delta-function ansatz
used in \cite{Ayala:2018ifo}.
For the BW case, the coefficient $k(M^2)$ takes the form
\begin{eqnarray}
k(M^2)\!=\! \frac{\int\limits_{4m_{\pi}^2}^{s_0}\! ds\,\left(\Delta_\rho(s)
	+\Delta_\omega(s)\right)m^2_\rho/s}
{	\!\!\int\limits_{4m_{\pi}^2}^{s_0}\!\!ds\left(\Delta_\rho(s)
	+\Delta_\omega(s)\right)\exp\left(\frac{m^2_\rho}{M^2}-\frac{s}{M^2}\right) }
	\, ,
\end{eqnarray}
where the BW spectral densities are described by means of
the mass $m_V$ and the width $\Gamma_V$ of
the included resonances of the rho and omega mesons ($V=\rho\, ,\omega$), i.e.,
\begin{eqnarray}
  \Delta_V(s)
\equiv
  \frac1{\pi}\frac{m_V\Gamma_V }{\left( m_V^2-s\right)^2
  +m_V^2 \Gamma_V^2} \ .
\end{eqnarray}
Replacing the BW model by a delta-function form $\Delta_V(s)\to\delta(s-m^2_V)$, one gets for the coefficient $k(M^2)$
\begin{eqnarray}
k(M^2) \to k_\delta(M^2) = 1 \, .
\end{eqnarray}
This reduces Eq.\ (\ref{eq:AppB-HV}) to the form given in \cite{Ayala:2018ifo}.
\end{appendix}


\end{document}